\documentclass[useAMS,usenatbib]{mn2e}

\usepackage{graphicx}
\usepackage{txfonts}
\usepackage{lscape}

\defcitealias{gusev2013a}{Paper~I}

\title[Star formation regions in spiral arms of NGC~628]
      {Parameters of the brightest star formation regions in the two 
       principal spiral arms of NGC~628}

\author[A.~S.~Gusev, O.~V.~Egorov and F.~Sakhibov]
       {A.~S.~Gusev,$^{1}$
        O.~V.~Egorov$^{1}$
        and F.~Sakhibov$^{2}$ \\
 $^{1}$ Sternberg Astronomical Institute, Lomonosov Moscow State University,
        Universitetsky pr. 13, 119992 Moscow, Russia, 
        {\sf gusev@sai.msu.ru} \\
 $^{2}$ University of Applied Sciences of Mittelhessen, Campus Friedberg,
        Department of Mathematics, Natural Sciences and Data Processing, \\
        Wilhelm-Leuschner-Strasse 13, 61169 Friedberg, Germany \\
        }

\date{Accepted 2013 October 10. Received 2013 October 10; in original 
form 2013 July 9}

\begin{document}

\maketitle

\begin{abstract}
We study photometric properties, chemical abundances and sizes of star 
formation regions in the two principal arms of the galaxy NGC~628 (M74). 
The {\it GALEX} ultraviolet, optical $UBVRI$, and H$\alpha$ surface photometry 
data are used, including those obtained with the 1.5-m telescope of the 
Maidanak Observatory. The thirty brightest star formation regions in 
ultraviolet light located in the spiral arms of NGC~628 are identified and 
studied. We find that the star formation regions in one (longer) arm 
are systematically brighter and larger than the regions in the other (shorter) 
arm. However, both luminosity and size distribution functions have 
approximately the same slopes for the samples of star formation regions in 
both arms. The star formation regions in the longer arm have a higher star 
formation rate density than the regions in the shorter arm. The regions in 
the shorter arm show higher N/O ratio at a higher oxygen abundance, but they 
have lower ultraviolet and H$\alpha$ luminosities. These findings can 
be explained if we assume that star formation regions in the shorter arm 
had higher star formation rate in the past, but now it is lower than for 
those in the opposite arm. Results of stellar evolutionary synthesis 
show that the brightest regions in the longer arm are slightly younger than 
the ones in the shorter arm ($3.5\pm2.2$~Myr versus $6.0\pm1.1$~Myr). Our 
results demonstrate that there is a difference in the inner structures and 
parameters of the interstellar medium between the spiral arms of NGC~628, one 
of which is long and hosts a regular chain of bright star formation complexes 
and the other, shorter one does not.
\end{abstract}

\begin{keywords}
H\,{\sc ii} regions -- galaxies: individual: NGC~628 (M74) -- 
galaxies: photometry -- ultraviolet: galaxies
\end{keywords}

\section{Introduction}

Star formation regions (H\,{\sc ii} regions) are associated with 
spiral arms of disc galaxies. Within spiral arms of grand design galaxies, 
star formation regions are often grouped into structures with sizes of about 
0.5~kpc, the star complexes 
\citep{elmegreen1977,efremov1978,efremov1979,elmegreen1996,efremov1998}. 
These complexes are the greatest coherent 
groupings of young stars. Such complexes are formed from H\,{\sc i}/H$_2$ 
superclouds \citep{elmegreen1983,efremov1989,efremov1995,elmegreen1994,
elmegreen2009,odekon2008,marcos2009}. The size/mass of the 
largest star formation regions that can appear in a galaxy is 
determined by the parameters of the interstellar medium, such as the 
gas density and pressure 
\citep*{elmegreen1997,kennicutt1998,billett2002,larsen2002}.

Occasionally, these star formation complexes are located along an arm at 
rather regular distances. \citet{elmegreen1983} found the spacing of complexes 
(H\,{\sc ii} regions) in studied galaxies to be within 1-4 kpc, and each 
string to consist, on average, of five H\,{\sc ii} regions.
\citet{elmegreen1983} and \citet{elmegreen1994} suggested that the 
gravitational or magneto-gravitational instability developing along the arm 
can explain this regularity. \citet{elmegreen1983} found that in two thirds 
of cases the regular strings of complexes are seen in one arm only. 
The well-known galaxy NGC~628 (M74) is the nearest object from the list of 
\citet{elmegreen1983} where the regular spacing of complexes are observed in 
one arm only. We believe that the study of the properties of such galaxies 
can help to better understand the nature of the regular chains of bright 
star formation complexes.

In a previous paper \citep[][hereafter \citetalias{gusev2013a}]{gusev2013a} 
we have studied photometric properties of spiral arms in NGC~628 and 
location of star formation regions inside these arms. Our results confirmed 
the conclusion of \citet{elmegreen1983}, that only one of the spiral arms in 
NGC~628 has the regular chain of bright star complexes. We also found 
that the characteristic separation between adjacent fainter star 
formation regions in both spiral arms of the galaxy is nearly 400~pc 
\citepalias{gusev2013a}. The main goal of this new research is to study 
differences between samples of bright star formation regions in two opposite 
spiral arms of NGC~628, and to understand why these samples differ from each 
other. Here, we consider photometric properties, chemical abundances and 
sizes of the brightest star formation regions in the two principal spiral 
arms of the grand-design galaxy NGC~628, based on our own observations in the 
$U$, $B$, $V$, $R$, $I$ passbands, and H$\alpha$ line, as well as the 
{\it Galaxy Evolution Explorer (GALEX)} far- and near-ultraviolet (FUV and 
NUV) data.

\begin{figure*}
\resizebox{0.90\hsize}{!}{\includegraphics[angle=000]{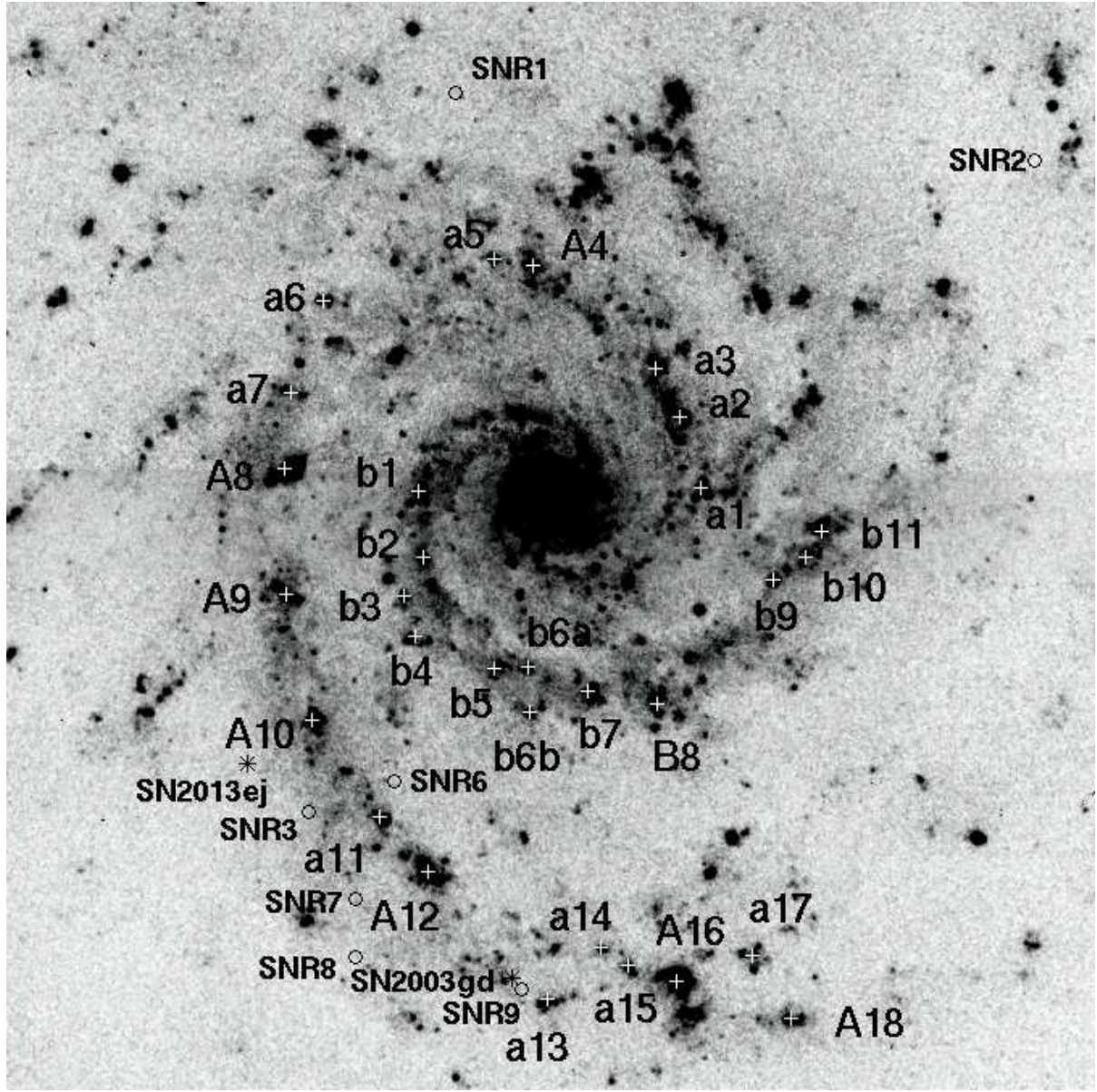}}
\caption{$U$ image of NGC~628 and positions of the galaxy's star formation 
regions, supernova remnants, and supernovae. The white crosses show 
positions of the studied regions. The ID numbers of the star formation regions 
from Table~\ref{table:sfrs_id} are indicated. The black circles show 
positions of the SNRs by \citet{sonbas2010}, the stars indicate the 
supernovae. Two supernova remnants (SNR~4 and SNR~5) and SN~2002ap are 
outside the image field. North is upward and east is to the left. The size of 
the image is $6.0\times6.0$ arcmin$^2$.
}
\label{figure:nfig5b}
\end{figure*}

NGC~628 is a nearby spiral galaxy viewed almost face-on 
(Fig.~\ref{figure:nfig5b}, Table~\ref{table:param}). It is an excellent 
example of a galaxy with regular strings of complexes which are seen in only 
one arm. \citet{elmegreen1983} found seven complexes (H\,{\sc ii} regions) 
with a characteristic separation of 1.6-1.7~kpc in one arm of the galaxy 
(Fig.~\ref{figure:nfig5b}, Table~\ref{table:sfrs_id}).

\begin{table}
\caption[]{\label{table:param}
Basic parameters of NGC~628.
}
\begin{center}
\begin{tabular}{ll} \hline \hline
Parameter                                & Value \\
\hline
Type                                     & SA(s)c \\
Total apparent $B$ magnitude ($B_t$)     & $9.70\pm0.26$ mag \\
Absolute $B$ magnitude ($M_B$)$^a$       & -20.72 mag \\
Inclination ($i$)                        & $7\degr\pm1\degr$ \\
Position angle (PA)                      & $25\degr$ \\
Heliocentric radial velocity ($v$)       & $659\pm1$ km\,s$^{-1}$ \\
Apparent corrected radius ($R_{25}$)$^b$ & $5.23\pm0.24$ arcmin \\
Apparent corrected radius ($R_{25}$)$^b$ & $10.96\pm0.51$ kpc \\
Distance ($d$)                           & 7.2 Mpc \\
Galactic absorption ($A(B)_{\rm Gal}$)   & 0.254 mag \\
Distance modulus ($m-M$)                 & 29.29 mag \\
\hline
\end{tabular}\\
\end{center}
\begin{flushleft}
$^a$ Absolute magnitude of the galaxy corrected for Galactic extinction and
inclination effect. \\
$^b$ Isophotal radius (25 mag\,arcsec$^{-2}$ in the $B$-band) corrected for 
Galactic extinction and absorption due to the inclination of NGC~628.
\end{flushleft}
\end{table}

NGC~628 is a galaxy that has experienced recent star formation episodes. 
\citet{hodge1976} identified 730 H\,{\sc ii} regions in the galaxy. 
\citet{sonbas2010} found nine supernova remnants (SNRs) in NGC~628 (see 
Fig.~\ref{figure:nfig5b}). Three supernovae (SN~2002ap, 2003gd, and 2013ej) 
have been observed in the galaxy since 2001.

NGC~628 is the largest member of a small group of galaxies. The group is 
centred around NGC~628 and the peculiar spiral NGC~660. NGC~628 is 
associated with several companions: UGC~1104, UGC~1171, UGC~1176 (DDO13), 
UGC~A20, KDG10, and dw0137+1541. Most of the companions are star-forming 
dwarf irregulars \citep{auld2006}. Two giant high velocity gas complexes 
($M {\rm (HI)}\sim(0.5-1)\times10^8M_\odot$) are located at $\sim10$~arcmin 
to the east and to the west from the galactic centre \citep{kamphuis1992}.

The distance to NGC~628 is still an open question. 
\citet*{sharina1996} obtained 7.2~Mpc based on their observations of the 
brightest supergiants in NGC~628. The same value was found by 
\citet*{vandyk2006}, who studied the optical curve of SN~2003gd. 
This value of the distance is in good agreement with the results of 
\citet*{mccall1985} and \citet{ivanov1992}, who studied 
global properties of NGC~628 and star complexes in it, respectively. An 
independent determination, based on observations of planetary nebulae, gave a 
value of 8.6~Mpc \citep{herrmann2008}. Alternative values, 9.3--9.9~Mpc, were 
obtained based on studies of SN~2003gd \citep{hendry2005,olivares2010} 
and the study of the gravitational stability of the gaseous disc 
of NGC~628 \citep{zasov1996}. A value close to 10~Mpc is favoured in 
studies of the NGC~628 Group by \citet{auld2006}. Following the 
recent studies of e.g. \citet{moustakas2010}, \citet{sonbas2010}, 
\citet{aniano2012} or \citet{berg2013} we use the value of the distance 
to NGC~628, obtained in \citet{sharina1996} and \citet{vandyk2006}. The 
adoption of an alternative value of the distance, 10~Mpc, will 
increase the luminosities and the linear distances (sizes) of the objects in 
NGC~628 by $\sim30\%$. However, this does not affect the main conclusions of 
our study, as we compare parameters of star formation regions in the spiral 
arms of the same galaxy.

The fundamental parameters of NGC~628 are presented in 
Table~\ref{table:param}. We use the position angle and 
the inclination of the galactic disc derived by \citet{sakhibov2004}. 
The morphological type and the Galactic absorption, $A(B)_{\rm Gal}$, 
are taken from the NED\footnote{http://ned.ipac.caltech.edu/} data base.
Other parameters are taken from the LEDA data 
base\footnote{http://leda.univ-lyon1.fr/} \citep{paturel2003}. We adopt the 
Hubble constant $H_0 = 75$ km\,s$^{-1}$Mpc$^{-1}$. With the assumed distance 
to NGC~628, we estimate a linear scale of 34.9~pc\,arcsec$^{-1}$.

The spiral arm with a regular string of complexes found by 
\citet{elmegreen1983} was named Arm~A, and the opposite arm was named Arm~B 
(Fig.~\ref{figure:nfig5b}). Arm~A is known as Arm~2 in 
\citet{kennicutt1976} and \citet{cornett1994} or South arm in 
\citet{rosales2011}.

\section{Observations and reduction}

The results of $UBVRI$ photometry of NGC~628 have already been 
published in \citet{bruevich2007}. H$\alpha$ spectrophotometric and 
{\it GALEX} ultraviolet photometric observations and data reduction for 
the galaxy have been described in \citetalias{gusev2013a}. Just a brief 
compilation is given for these observations and data reduction.

The photometric and spectrophotometric CCD observations were obtained in 2002 
($UBVRI$) and 2006 (H$\alpha$) with the 1.5-m telescope of the Maidanak 
Observatory (Institute of Astronomy of the Academy of Sciences of Uzbekistan). 
The focal length of the telescope is 12~m.  A detailed description 
of the telescope and the CCD camera can be found in \citet{artamonov2010}. 
The images have a pixel scale of 0.267~arcsec\,pixel$^{-1}$. The seeing 
during the observations was 0.7--1.1~arcsec.

Ultraviolet {\it GALEX} FUV and NUV reduced FITS-images of NGC~628 were 
downloaded from Barbara A. Miculski archive for space telescopes 
(galex.stsci.edu; source GI3\_050001\_NGC628). The observations were made in 
2007. The description of the {\it GALEX} mission and basic parameters of 
passbands are presented in \citet{morrissey2005}. The image resolution is 
equal to 4.5~arcsec for FUV and 6.0~arcsec for NUV.

The reduction of the photometric and spectrophotometric data was carried out 
using standard techniques, with the European Southern Observatory Munich 
Image Data Analysis System\footnote{http://www.eso.org/sci/software/esomidas/} 
({\sc eso-midas}) \citep{banse1983,grosbol1990}. The main photometric and 
spectrophotometric image reduction stages were described in detail in 
\citet{bruevich2007} and \citetalias{gusev2013a}.

We corrected all data for Galactic absorption using the calibration of 
\citet{schlafly2011}; these values are indicated by the '0' subscript. We 
used the resulting ratio of the extinction in the {\it GALEX} bands to the 
color excess $A_{\rm FUV}/E(B-V)=8.24$ and $A_{\rm NUV}/E(B-V)=8.2$ 
\citep{wyder2007}.

To find and select star formation regions, we measured the magnitudes of 
the brightest regions in the spiral arms of the galaxy. The 
photometry was made using round apertures, and the light from the 
surrounding background was subtracted from the light coming from the area 
occupied by the star formation region. The technique of star formation 
regions photometry is described in more detail in \citet{gusev2003} and 
\citet{bruevich2007}.

\begin{table}
\caption[]{\label{table:sfrs_id}
Offsets and identification of star formation regions in the arms.
}
\begin{center}
\begin{tabular}{cccccccc} \hline \hline
Re-  & ID  & N--S$^a$ & E--W$^a$ & ID1$^b$ & ID2$^c$ & ID3$^d$ & ID4 \\
gion &     & (arcsec) & (arcsec) &         &         &         &     \\
(1)  & (2) & (3)      & (4)      & (5)     & (6)     & (7)     & (8) \\
\hline
 1 & a1  &   +2.13 &  -49.61 & --- & 12  & 100 & ---    \\
 2 & a2  &  +25.60 &  -42.68 & --- & 13  & --- & ---    \\
 3 & a3  &  +41.60 &  -34.68 & --- & 14  & 114 & ---    \\
 4 & A4  &  +75.73 &   +5.86 & --- & 20  &  11 & 1$^e$  \\
 5 & a5  &  +77.87 &  +18.66 & --- & --- &  12 & ---    \\
 6 & a6  &  +64.00 &  +75.19 & --- & 23  &  29 & ---    \\
 7 & a7  &  +33.60 &  +85.86 & --- & --- & --- & ---    \\
 8 & A8  &   +8.53 &  +87.99 & A1  & 60+ &  30 & ---    \\
   &     &         &         &     & 61  &     &        \\
 9 & A9  &  -33.07 &  +87.46 & A2  & 65  &  53 & ---    \\
10 & A10 &  -74.67 &  +78.93 & A3  & 80  &  61 & 4$^e$  \\
11 & a11 & -106.67 &  +56.53 & A4  & 82  &  66 & ---    \\
12 & A12 & -124.80 &  +40.53 & A5  & 84  & 68+ & 4$^f$  \\
   &     &         &         &     &     & 69  &        \\
13 & a13 & -166.94 &   +1.06 & --- & 91  & --- & ---    \\
14 & a14 & -149.87 &  -16.54 & --- & --- &  82 & ---    \\
15 & a15 & -155.74 &  -25.61 & --- & 93  &  84 & ---    \\
16 & A16 & -161.07 &  -41.61 & A6  & 94  &  83 & 5$^f$  \\
   &     &         &         &     &     &     & 6$^e$  \\
17 & a17 & -152.54 &  -66.68 & --- & --- & 85+ & ---    \\
   &     &         &         &     &     & 86  &        \\
18 & A18 & -173.34 &  -79.48 & A7  & --- & --- & ---    \\
19 & b1  &   +1.06 &  +43.73 & --- &  6  &  25 & 6$^g$  \\
20 & b2  &  -20.80 &  +42.13 & --- &  7  &  50 & ---    \\
21 & b3  &  -33.60 &  +48.53 & --- &  9  &  52 & ---    \\
22 & b4  &  -46.94 &  +44.79 & --- & 67  & --- & ---    \\
23 & b5  &  -57.60 &  +18.66 & --- & --- & --- & ---    \\
24 & b6a &  -57.07 &   +7.46 & --- & 68  & --- & ---    \\
25 & b6b &  -72.00 &   +7.19 & --- & 69  & 63+ & ---    \\
   &     &         &         &     &     & 64  &        \\
26 & b7  &  -65.07 &  -12.27 & --- & 71  & --- & ---    \\
27 & B8  &  -69.34 &  -35.21 & --- & 58+ & 77+ & 3$^g$  \\
   &     &         &         &     & 72  & 78  &        \\
28 & b9  &  -28.27 &  -73.61 & --- & 56  &  89 & 2$^e$  \\
   &     &         &         &     &     &     & 2$^g$  \\
   &     &         &         &     &     &     & A$^h$  \\
29 & b10 &  -20.80 &  -84.28 & --- & --- &  91 & ---    \\
30 & b11 &  -12.27 &  -89.61 & --- & --- &  92 & 2$^g$  \\
\hline
\end{tabular}
\end{center}
\begin{flushleft}
$^a$ Offsets from the galactic centre, positive to the north and west. \\
$^b$ ID by \citet{elmegreen1983}. \\
$^c$ ID by \citet{rosales2011}. \\
$^d$ ID by \citet{belley1992}. \\
$^e$ Ordinal numbers from Table~5 of \citet{berg2013}. \\
$^f$ ID by \citet*{bresolin1999}; the list of \citet{bresolin1999} coincides 
with the list of \citet{mccall1985}. \\ 
$^g$ ID by \citet{gusev2012}. \\
$^h$ ID by \citet*{ferguson1998}.
\end{flushleft}
\end{table}

\begin{figure}
\vspace{7.0mm}
\resizebox{0.98\hsize}{!}{\includegraphics[angle=000]{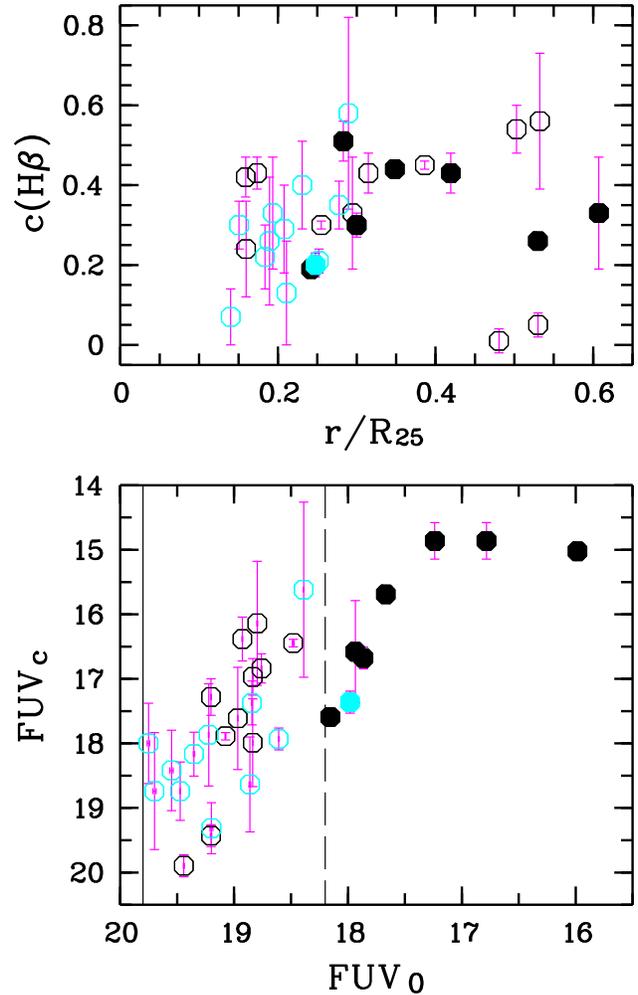}}
\caption{Top panel: radial distribution of interstellar absorption in 
star formation regions. Bottom panel: comparison between uncorrected and 
corrected for interstellar absorption magnitudes FUV of the regions. Lower 
limits of magnitude FUV$_0$ for bright complexes (dashed line) and 
star formation regions (solid line) are shown. The black circles denote the 
regions in Arm~A, the grey circles show the objects in Arm~B. The 
filled circles are bright complexes, and the open circles are fainter star 
formation regions. The magnitudes error bars are shown.
}
\label{figure:nfig8}
\end{figure}

As a result, we selected 30 star formation regions having a total magnitude 
FUV$_0 < 19.8$~mag (Fig.~\ref{figure:nfig5b}). The objects were divided into 
bright complexes and fainter star formation regions. Eight complexes brighter 
than 18.2~mag in FUV were selected as 'bright complexes', other 22 objects 
were named 'star formation regions'. We cut the list of bright complexes on 
the seventh brightest complex in Arm~A; it coincides with the bright 
H\,{\sc ii} regions list of \citet{elmegreen1983} but for one exception, 
our complex A4 is brighter in FUV than the star formation region a11 
(Tables~\ref{table:sfrs_id}, \ref{table:sfrs_photcor}, 
Fig.~\ref{figure:nfig5b}). Among regions fainter than 19.8~mag in FUV, we 
found a large number of diffuse objects without a strong H$\alpha$ emission. 
Such objects are rather close groups of stellar associations. They have 
not been investigated in this study. We will show below, that the variation 
of the limits of brightness does not affect our conclusions in principle.

Spatial location of the star formation regions are shown in 
Fig.~\ref{figure:nfig5b}. Galactocentric coordinates and identification 
data of the star formation regions in the arms of NGC~628 are presented in 
Table~\ref{table:sfrs_id}. Ordinal and identification numbers of the star 
formation regions are given in columns (1) and (2), respectively. Offsets 
from the galactic centre are presented in columns (3) and (4), respectively. 
Identification numbers of the objects by \citet{elmegreen1983} are shown in 
column (5). Most of the selected star formation regions were studied 
earlier based on spectroscopic and spectrophotometric observations 
\citep{mccall1985,belley1992,ferguson1998,bresolin1999,rosales2011,gusev2012,
berg2013}. The cross identification data for the complexes are also 
presented in Table~\ref{table:sfrs_id}. Identification numbers of 
\citet{rosales2011} are given in column (6), numbers of \citet{belley1992} 
are presented in column (7), and identification numbers of \citet{berg2013}, 
\citet{bresolin1999}, \citet{ferguson1998}, and \citet{gusev2012} are given 
in column (8).

We used a letter-number identification for the star formation regions, the 
letter 'a' is used for the regions in Arm~A, and the letter 'b' is used for 
the objects in Arm~B. Bright complexes are marked by a capital letter, and 
fainter star formation regions are marked by a small one. Sequential 
numbering is used for the regions in every arm, based on their longitudinal 
displacements along the arm (Fig.~\ref{figure:nfig5b}, 
Table~\ref{table:sfrs_id}). Two star formation regions in Arm~B are located 
at the same longitudinal displacement along the spiral arm, they were named 
'b6a' and 'b6b'.

\begin{landscape}
\begin{table}
\caption[]{\label{table:sfrs_photcor}
Photometric parameters, diameters and galactocentric distances of star 
formation regions.
}
\begin{center}
\begin{tabular}{cccccccccccccccc} \hline \hline
ID  & $r/R_{25}^a$ & $r^b$ & $A^c$    & FUV$_0$ & $M{\rm (FUV)}_{\rm c}$ & (FUV- & 
(NUV- & $(U-B)_{\rm c}$ & $(B-V)_{\rm c}$ & $(V-R)_{\rm c}$ & 
$(V-I)_{\rm c}$ & $\log I$(H$\alpha$) & $c$(H$\beta$) & $d$ & $l^d$ \\
    &      &       &       &       &       & ${\rm NUV)_c}$ & 
$U)_{\rm c}$  &           &           &           &           & 
([erg \,s$^{-1}$ &     &     &     \\
    &              & (kpc) & (arcsec) & (mag)   & (mag)      & (mag)    &
(mag)       & (mag)     & (mag)     & (mag)     & (mag)     & 
cm$^{-2}$])      &     & (pc) & (kpc)  \\
\hline
a1  & 0.159 & 1.74 & 12.8 & 18.84 & -12.32$\pm$0.28 &  0.28$\pm$0.40 &  0.15$\pm$0.33 & -1.01$\pm$0.22 & -0.56$\pm$0.19 &  0.56$\pm$0.16 & -0.75$\pm$0.30 & -12.94$\pm$0.03 & 0.42$\pm$0.05$^e$ & 365$\pm$25 & 0.85  (a2) \\
a2  & 0.160 & 1.75 &  9.1 & 18.84 & -11.30$\pm$0.68 &  0.41$\pm$0.96 &  0.85$\pm$0.78 & -0.69$\pm$0.52 & -0.09$\pm$0.43 &  0.27$\pm$0.34 &  0.07$\pm$0.31 & -12.99$\pm$0.08 & 0.24$\pm$0.12$^e$ & 325$\pm$35 & 0.62  (a3) \\
a3  & 0.174 & 1.90 & 11.7 & 18.76 & -12.45$\pm$0.23 &  0.35$\pm$0.32 &  0.44$\pm$0.26 & -0.86$\pm$0.17 & -0.22$\pm$0.15 &  0.19$\pm$0.12 & -0.05$\pm$0.12 & -12.79$\pm$0.03 & 0.43$\pm$0.04$^e$ & 340$\pm$25 & 0.62  (a2) \\
A4  & 0.242 & 2.65 & 14.9 & 18.16 & -11.70$\pm$0.11 &  0.30$\pm$0.16 &  0.80$\pm$0.13 & -0.78$\pm$0.09 & -0.07$\pm$0.08 &  0.39$\pm$0.07 & -0.11$\pm$0.09 & -12.74$\pm$0.01 & 0.19$\pm$0.02$^e$ & 450$\pm$45 & 0.46  (a5) \\
a5  & 0.255 & 2.80 &  9.1 & 19.08 & -11.40$\pm$0.06 &  0.24$\pm$0.08 &  0.85$\pm$0.07 & -0.86$\pm$0.05 & -0.22$\pm$0.05 &  0.45$\pm$0.04 & -0.09$\pm$0.05 & -12.95$\pm$0.01 & 0.30$\pm$0.01$^f$ & 280$\pm$15 & 0.46  (A4) \\
a6  & 0.315 & 3.45 &  8.5 & 19.20 & -12.01$\pm$0.28 &  0.20$\pm$0.40 &  0.43$\pm$0.33 & -1.00$\pm$0.22 & -0.31$\pm$0.18 &  0.27$\pm$0.14 & -0.65$\pm$0.14 & -13.02$\pm$0.03 & 0.43$\pm$0.05$^e$ & 275$\pm$15 & 1.13  (a7) \\
a7  & 0.295 & 3.23 & 11.7 & 18.97 & -11.68$\pm$0.79 &  0.28$\pm$1.12 &  0.48$\pm$0.91 & -0.51$\pm$0.61 & -0.30$\pm$0.50 & -0.10$\pm$0.39 &  0.84$\pm$0.35 & -13.52$\pm$0.09 & 0.33$\pm$0.14$^g$ & 290$\pm$45 & 0.88  (A8) \\
A8  & 0.283 & 3.10 & 19.7 & 17.24 & -14.43$\pm$0.28 &  0.25$\pm$0.40 & -0.07$\pm$0.33 & -0.75$\pm$0.22 & -0.32$\pm$0.18 &  0.10$\pm$0.14 & -0.36$\pm$0.13 & -12.46$\pm$0.03 & 0.51$\pm$0.05$^e$ & 565$\pm$65 & 0.88  (a7) \\
A9  & 0.300 & 3.29 & 16.0 & 17.86 & -12.61$\pm$0.17 &  0.25$\pm$0.24 &  0.59$\pm$0.20 & -0.56$\pm$0.13 & -0.10$\pm$0.11 &  0.21$\pm$0.08 & -0.03$\pm$0.08 & -12.78$\pm$0.02 & 0.30$\pm$0.03$^e$ & 445$\pm$30 & 1.45  (A8) \\
A10 & 0.348 & 3.82 & 13.9 & 17.67 & -13.60$\pm$0.11 &  0.25$\pm$0.16 &  0.10$\pm$0.13 & -0.94$\pm$0.09 & -0.25$\pm$0.07 &  0.23$\pm$0.06 & -0.10$\pm$0.05 & -12.56$\pm$0.01 & 0.44$\pm$0.02$^e$ & 495$\pm$40 & 1.36 (a11) \\
a11 & 0.386 & 4.23 & 13.9 & 18.48 & -12.84$\pm$0.06 &  0.24$\pm$0.08 &  0.26$\pm$0.07 & -0.80$\pm$0.04 & -0.27$\pm$0.04 &  0.28$\pm$0.03 & -0.23$\pm$0.03 & -12.72$\pm$0.01 & 0.45$\pm$0.01$^e$ & 405$\pm$35 & 0.84 (A12) \\
A12 & 0.419 & 4.60 & 20.8 & 16.78 & -14.43$\pm$0.28 &  0.22$\pm$0.40 &  0.16$\pm$0.33 & -0.85$\pm$0.22 & -0.16$\pm$0.18 &  0.28$\pm$0.14 & -0.09$\pm$0.13 & -12.16$\pm$0.03 & 0.43$\pm$0.05$^e$ & 570$\pm$20 & 0.84 (a11) \\
a13 & 0.532 & 5.83 & 10.7 & 18.80 & -13.15$\pm$0.96 &  0.20$\pm$1.36 & -0.08$\pm$1.11 & -0.92$\pm$0.74 & -0.33$\pm$0.60 &  0.02$\pm$0.47 &  0.09$\pm$0.43 & -13.11$\pm$0.11 & 0.56$\pm$0.17$^e$ & 335$\pm$20 & 0.86 (a14) \\
a14 & 0.481 & 5.27 &  7.5 & 19.44 &  -9.39$\pm$0.17 &  0.00$\pm$0.24 &  1.31$\pm$0.20 & -0.70$\pm$0.13 & -0.17$\pm$0.11 &  0.33$\pm$0.09 &  0.39$\pm$0.08 & -13.68$\pm$0.02 & 0.01$\pm$0.03$^f$ & 260$\pm$15 & 0.38 (a15) \\
a15 & 0.503 & 5.51 & 10.7 & 18.93 & -12.91$\pm$0.34 &  0.15$\pm$0.48 &  0.03$\pm$0.39 & -0.82$\pm$0.26 & -0.29$\pm$0.21 &  0.39$\pm$0.17 &  0.19$\pm$0.15 & -12.83$\pm$0.04 & 0.54$\pm$0.06$^e$ & 280$\pm$40 & 0.38 (a14) \\
A16 & 0.530 & 5.81 & 28.3 & 15.98 & -14.27$\pm$0.06 &  0.18$\pm$0.08 &  0.49$\pm$0.07 & -0.74$\pm$0.04 & -0.03$\pm$0.04 &  0.19$\pm$0.03 &  0.09$\pm$0.03 & -12.03$\pm$0.01 & 0.26$\pm$0.01$^e$ & 795$\pm$40 & 0.59 (a15) \\
a17 & 0.530 & 5.81 & 14.4 & 19.20 &  -9.86$\pm$0.17 &  0.14$\pm$0.24 &  1.37$\pm$0.20 & -0.66$\pm$0.13 &  0.05$\pm$0.11 &  0.30$\pm$0.09 &  0.42$\pm$0.08 & -13.44$\pm$0.02 & 0.05$\pm$0.03$^f$ & 350$\pm$20 & 0.85 (A18) \\
A18 & 0.607 & 6.66 & 13.9 & 17.94 & -12.71$\pm$0.79 &  0.12$\pm$1.12 &  0.28$\pm$0.91 & -0.81$\pm$0.61 & -0.09$\pm$0.50 &  0.06$\pm$0.39 & -0.10$\pm$0.35 & -12.92$\pm$0.09 & 0.33$\pm$0.14$^g$ & 410$\pm$30 & 0.85 (a17) \\
b1  & 0.140 & 1.54 & 12.8 & 19.20 &  -9.98$\pm$0.40 &  0.32$\pm$0.56 &  1.33$\pm$0.46 & -0.53$\pm$0.31 &  0.05$\pm$0.26 &  0.14$\pm$0.22 &  0.12$\pm$0.22 & -13.42$\pm$0.05 & 0.07$\pm$0.07$^e$ & 290$\pm$30 & 0.77  (b2) \\
b2  & 0.151 & 1.65 &  9.1 & 19.35 & -11.12$\pm$0.34 &  0.40$\pm$0.48 &  0.83$\pm$0.39 & -0.62$\pm$0.26 &  0.03$\pm$0.21 &  0.19$\pm$0.17 & -0.04$\pm$0.16 & -13.15$\pm$0.04 & 0.30$\pm$0.06$^e$ & 250$\pm$15 & 0.50  (b3) \\
b3  & 0.189 & 2.08 &  9.1 & 19.70 & -10.55$\pm$0.91 &  0.31$\pm$1.28 &  0.98$\pm$1.04 & -1.00$\pm$0.69 & -0.42$\pm$0.58 &  0.24$\pm$0.47 & -0.47$\pm$0.47 & -13.34$\pm$0.11 & 0.26$\pm$0.16$^e$ & 235$\pm$20 & 0.48  (b4) \\
b4  & 0.208 & 2.28 & 11.7 & 19.55 & -10.87$\pm$0.62 &  0.29$\pm$0.88 &  0.86$\pm$0.72 & -0.52$\pm$0.48 & -0.44$\pm$0.39 & -0.07$\pm$0.32 & -0.25$\pm$0.30 & -14.06$\pm$0.08 & 0.29$\pm$0.11$^e$ & 270$\pm$30 & 0.48  (b3) \\
b5  & 0.193 & 2.12 & 10.1 & 19.22 & -11.42$\pm$0.79 &  0.33$\pm$1.12 &  0.69$\pm$0.91 & -0.35$\pm$0.61 & -0.12$\pm$0.50 &  0.11$\pm$0.39 &  0.00$\pm$0.36 & -13.53$\pm$0.09 & 0.33$\pm$0.14$^g$ & 245$\pm$35 & 0.39 (b6a) \\
b6a & 0.184 & 2.01 &  8.0 & 19.47 & -10.55$\pm$0.45 &  0.31$\pm$0.64 &  0.95$\pm$0.52 & -0.54$\pm$0.35 & -0.10$\pm$0.29 &  0.31$\pm$0.24 &  0.38$\pm$0.22 & -13.34$\pm$0.05 & 0.22$\pm$0.08$^e$ & 240$\pm$30 & 0.39  (b5) \\
b6b & 0.231 & 2.53 & 11.2 & 19.75 & -11.29$\pm$0.62 &  0.26$\pm$0.88 &  0.76$\pm$0.72 & -0.87$\pm$0.48 & -0.09$\pm$0.39 &  0.26$\pm$0.31 & -0.07$\pm$0.29 & -13.14$\pm$0.07 & 0.40$\pm$0.11$^e$ & 225$\pm$35 & 0.52 (b6a) \\
b7  & 0.211 & 2.31 & 13.3 & 18.86 & -10.65$\pm$0.74 &  0.28$\pm$1.04 &  1.26$\pm$0.85 & -0.44$\pm$0.56 &  0.04$\pm$0.46 &  0.06$\pm$0.36 &  0.24$\pm$0.33 & -13.55$\pm$0.09 & 0.13$\pm$0.13$^e$ & 330$\pm$25 & 0.73 (b6b) \\
B8  & 0.248 & 2.71 & 18.7 & 17.98 & -11.93$\pm$0.17 &  0.32$\pm$0.24 &  0.91$\pm$0.20 & -0.70$\pm$0.13 & -0.14$\pm$0.11 &  0.24$\pm$0.10 & -0.02$\pm$0.11 & -12.72$\pm$0.02 & 0.20$\pm$0.03$^e$ & 445$\pm$65 & 0.82  (b7) \\
b9  & 0.252 & 2.76 & 12.8 & 18.61 & -11.36$\pm$0.17 &  0.37$\pm$0.24 &  1.03$\pm$0.20 & -0.60$\pm$0.13 &  0.22$\pm$0.11 &  0.45$\pm$0.09 &  0.14$\pm$0.08 & -12.67$\pm$0.02 & 0.21$\pm$0.03$^e$ & 305$\pm$30 & 0.46 (b10) \\
b10 & 0.278 & 3.04 & 12.3 & 18.84 & -11.92$\pm$0.34 &  0.33$\pm$0.48 &  0.62$\pm$0.39 & -0.38$\pm$0.26 & -0.07$\pm$0.21 &  0.09$\pm$0.17 & -0.09$\pm$0.16 & -13.27$\pm$0.04 & 0.35$\pm$0.06$^f$ & 325$\pm$45 & 0.35 (b11) \\
b11 & 0.290 & 3.17 & 12.8 & 18.39 & -13.67$\pm$1.36 &  0.28$\pm$1.92 &  0.05$\pm$1.56 & -0.64$\pm$1.04 & -0.21$\pm$0.85 &  0.11$\pm$0.67 &  0.05$\pm$0.60 & -12.85$\pm$0.16 & 0.58$\pm$0.24$^h$ & 400$\pm$25 & 0.35 (b10) \\
\hline
\end{tabular}
\end{center}
\begin{flushleft}
$^a$ Deprojected galactocentric distance normalized to the disc isophotal 
radius $R_{25}$. \\ $^b$ Deprojected galactocentric distance. \\ 
$^c$ Diameter of the aperture. \\ $^d$ Distance to the nearest neighbour, 
the ID numbers of the nearest star formation regions/complexes are shown in
the brackets. \\ $^e$ \citet{rosales2011}. \\ $^f$ \citet{belley1992}. \\ 
$^g$ Mean for the regions from our list by data of \citet{rosales2011}. \\ 
$^h$ \citet{gusev2012}.
\end{flushleft}
\end{table}
\end{landscape}

\section{Star formation regions in the arms}

\subsection{Photometric parameters of star formation regions}

Results of photometric observations of the star formation regions using 
a round aperture are given in Table~\ref{table:sfrs_photcor}. 
Magnitudes and H$\alpha$ fluxes in this table are corrected for interstellar 
absorption. Taking into account the interstellar absorption is extremely 
important for obtaining real luminosities and colour indices and study 
the physical parameters of star formation regions. We used a 
logarithmic extinction coefficient, $c$(H$\beta$), obtained from 
spectroscopic and spectrophotometric observations, to correct the 
photometric data for interstellar absorption in the regions 
(see Table~\ref{table:sfrs_id}). The reddening function of 
\citet{cardelli1989} was adopted, assuming $R \equiv A_V/E(B-V) = 3.1$, for 
correction of fluxes in optical bands, and data of \citet{wyder2007} were 
used for correction of fluxes in ultraviolet bands.

The most complete contemporary study of spectral parameters of 
H\,{\sc ii} regions was carried out by \citet{rosales2011}, who used data of 
integral field spectroscopy of NGC~628. Estimations of logarithmic extinction 
coefficients for most of objects, studied here, were derived by 
\citet{rosales2011}. These estimations are used in the present paper. For 
other objects we accept estimations of $c$(H$\beta$) from \citet{belley1992} 
and \citet{gusev2012}. Note that the accuracy of $c$(H$\beta$) 
estimations derived from spectrophotometric observations of 
\citet{belley1992} is lower than the ones based on the spectroscopy. 
There are no spectroscopic or spectrophotometric observations for three 
objects (a7, A18, b5). For these regions, we use $c$(H$\beta$) = $0.33\pm0.14$ 
as the mean value for the complexes in our list. Adopted $c$(H$\beta$) are 
presented in Table~\ref{table:sfrs_photcor}.

Colour indices and H$\alpha$ fluxes, corrected for interstellar absorption 
absolute magnitudes, are presented in Table~\ref{table:sfrs_photcor}. 
The interstellar absorption, calculated using $c$(H$\beta$), includes the
Galactic extinction, the internal extinction due to the interstellar medium 
within NGC~628, and the intergalactic extinction due to the intergalactic 
medium between the Milky Way and NGC~628. These values are indicated by 'c' 
subscript.

The main contribution to the inaccuracy of magnitudes, corrected for 
interstellar absorption, is related to the uncertainty in the extinction 
coefficient, especially in the short wavelength bands. Obviously, the results 
of photometry can be used only for qualitative comparison of physical 
parameters of the star formation regions in Arms~A and B.

After correction for interstellar absorption, as we can see from 
Fig.~\ref{figure:nfig8}, some 'bright' complexes become fainter than some 
'faint' star formation regions and vice versa. However, it does not 
affect the following conclusions. Below we study samples 
of brightest star formation regions in Arms~A and B without a division of 
objects in 'bright complexes' and 'star formation regions'.

The value of interstellar absorption in star formation regions of Arm~A 
and Arm~B is approximately the same: 
$\langle c$(H$\beta$)$\rangle = 0.35\pm0.15$ versus $0.28\pm0.13$. It 
does not depend on the galactocentic distance for regions in Arm~A, 
$\langle c$(H$\beta$)$\rangle = 0.35\pm0.10$ for regions with 
$r/R_{25}<0.32$ and $0.34\pm0.20$ for regions with $r/R_{25}>0.32$ 
(Fig.~\ref{figure:nfig8}). Note a large variation of $c$(H$\beta$) for 
objects at the end of Arm~A (Fig.~\ref{figure:nfig8}).

Thus, photometric data, corrected for interstellar absorption, support that 
young stellar objects (both complexes and star formation regions) in 
Arm~A are systematically brighter than the ones in Arm~B of NGC~628. 
Below we will discuss the physical reasons of such differences.

\begin{figure}
\vspace{7.5mm}
\resizebox{0.90\hsize}{!}{\includegraphics[angle=000]{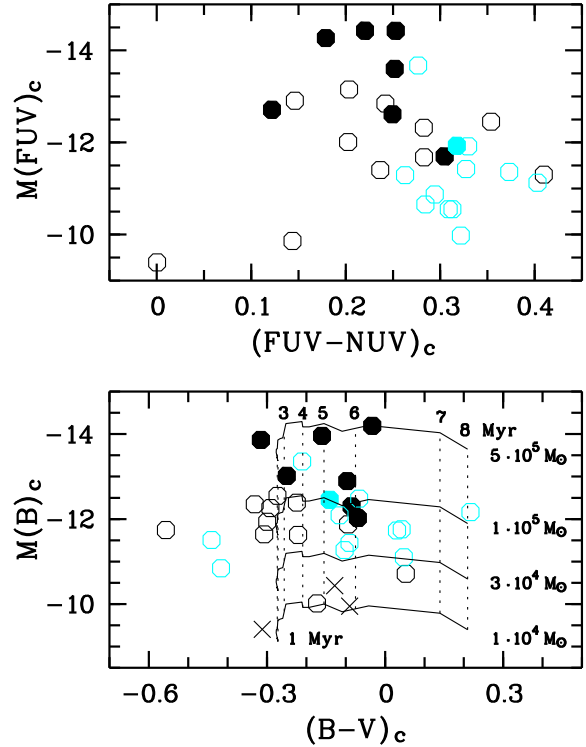}}
\caption{
$M({\rm FUV})_{\rm c}$~vs.~(FUV-NUV)$_{\rm c}$ (top) and 
$M(B)_{\rm c}$~vs.~$(B-V)_{\rm c}$ (bottom) colour-magnitude diagrams for 
regions in arms of NGC~628. Evolutionary tracks of synthetically aged stellar 
systems of different masses (solid curves) are shown. The dotted lines are 
isochrones of synthetic stellar systems. The diagonal crosses show open star 
clusters in the Milky Way by data of \citet{kharchenko2009}. Other symbols 
are the same as in Fig.~\ref{figure:nfig8}. See the text for details.
}
\label{figure:nfig9}
\end{figure}

\begin{figure}
\vspace{7.0mm}
\resizebox{1.00\hsize}{!}{\includegraphics[angle=000]{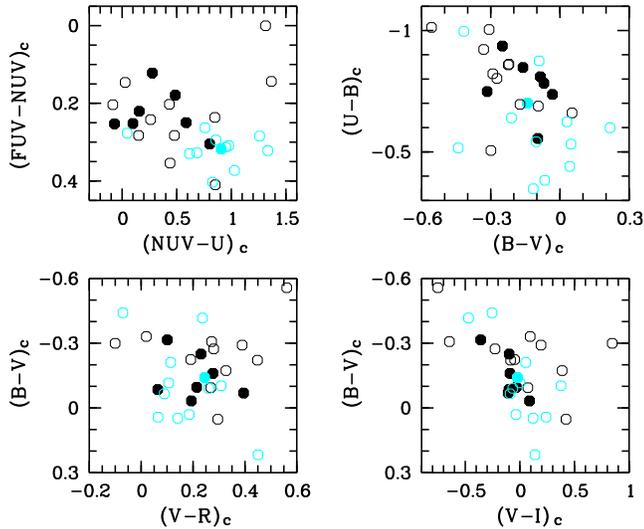}}
\caption{Two-colour diagrams for the star formation regions in the arms of 
NGC~628. Symbols are the same as in Fig.~\ref{figure:nfig8}.
}
\label{figure:nfig10}
\end{figure}

Star formation regions in Arm~A are bluer than the ones in Arm~B 
(Figs.~\ref{figure:nfig9}, \ref{figure:nfig10}). Differences between colour 
properties of regions in the arms decrease toward long wavelength 
passbands. Two relatively well defined groups of regions appear in the 
ultraviolet colour-magnitude diagram, 
(FUV-NUV)$_{\rm c}$~vs.~$({\rm NUV}-U)_{\rm c}$ and 
$(U-B)_{\rm c}$~vs.~$(B-V)_{\rm c}$ two-colour diagrams, and are mixed in the 
$(B-V)_{\rm c}$~vs.~$(V-R)_{\rm c}$ and $(B-V)_{\rm c}$~vs.~$(V-I)_{\rm c}$ 
two-colour diagrams (Figs.~\ref{figure:nfig9}, \ref{figure:nfig10}).

In Fig.~\ref{figure:nfig9} (bottom) we compare the observed colour-magnitude 
relations obtained in $B$ and $V$ passbands for studied objects with the 
prediction of standard SSP-models (Stellar population synthesis model 
predictions). A number of SSP-models have been constructed during the last 
decade. They are widely used for modelling both star clusters and galactic 
populations. Photometric properties of model clusters are defined by the 
implemented grid of isochrones. Here we use the grid provided by the 
Padova group \citep{bertelli1994,girardi2000,marigo2007,marigo2008} via the 
online server CMD\footnote{http://stev.oapd.inaf.it/cgi-bin/cmd}. The 
latest Padova models (version 2.5), described in \citet{bressan2012}, 
are computed for the narrower interval of initial masses ranges from 
0.1~$M_\odot$ to 12~$M_\odot$. For our purposes we need the interval of 
initial masses ranges up to 100~$M_\odot$. That is the reason, why we used the 
prior sets of stellar evolutionary tracks (version 2.3), described in 
\citet{marigo2008} and computed for the wide interval of initial masses 
ranges from 0.15~$M_\odot$ to 100~$M_\odot$.

We used a metallicity grid with $Z = 0.012$ which is close to the mean 
chemical abundance of H\,{\sc ii} regions in NGC~628, and retrieved the 
passbands $B$, $V$ an age range $\log t = 6.0 - 10.2$ and a step of 0.05 
in $\log t$. Calculations of integrated $L_B$ and $L_V$ fluxes are performed 
for the case of a continuous populated IMF and simultaneous star formation, 
according to the method described in the paper by \citet{piskunov2009}. We 
computed a number of models for the different mass values of star clusters 
from $10^4 M_\odot$ up to $3.5\times10^5 M_\odot$. We assumed a Salpeter 
value of the slope $\alpha = -2.35$ and low mass limit $m_l=0.1 M_\odot$ of 
the IMF. The upper limit $m_l=100 M_\odot$ of the IMF is limited by the used 
evolutionary grid. Fig.~\ref{figure:nfig9} shows four evolutionary tracks 
computed for different masses of the model and for the age interval from 
1~Myr up to 8~Myr. These parameters were chosen to provide a fit of the 
colour distribution on the colour-magnitude diagram.

About $80\%$ of the luminosity of star formation regions in the $B$ band 
is provided by high mass stars ($m>4M_\odot$). The $B-V$ colour indices of these 
massive stars are approximately similar within the main sequence at fixed age. 
One can see that isochrones of synthetic clusters of different masses in 
the colour-magnitude diagram (Fig.~\ref{figure:nfig9}, bottom) are 
perpendicular to the $B-V$ axis. The young massive regions studied here may 
be made of several star clusters produced in a single episode of star 
formation and having identical ages and thereby have identical $B-V$ colour 
indices of the brightest stars. It means that the multiple structure of 
unresolved star forming regions does not influence the integrated $B-V$ 
colour indices of unresolved star formation regions. In case of $U-B$ colours, 
the mass dispersion of individual clusters embedded into the unresolved star 
formation regions leads to slight reddening of the integrated $U-B$ 
colour indices and thereby to older ages.

Since we use the $B$ luminosities and integrated $B-V$ colours of unresolved 
star formation regions in the colour-magnitude diagram, we can assign 
parameters of the model of the single massive cluster to the unresolved 
multiple star clusters.

Fig.~\ref{figure:nfig9} shows that all studied objects are younger than 
8~Myr. The figure shows also that typical mass interval of studied star 
formation regions are within the range from $1\times10^4 M_\odot$ up to 
$\approx5\times10^5 M_\odot$. The lower limit of the mass interval overlaps 
with the upper mass limit of open star clusters (OSCs) in the Milky Way. 
The three brightest complexes in Arm~A (A8, A12 and A16) have 
approximately the same luminosity. The synthetic evolutionary tracks show 
that the mass of these complexes is $5\times10^5 M_\odot$ 
(Fig.~\ref{figure:nfig9}). The upper limit of the mass interval is close to 
masses of young massive star clusters in nearby galaxies \citep{larsen2011}.

Results of stellar evolutionary synthesis show that star formation regions in 
Arm~A are slightly younger than the ones in Arm~B. Excluding the 
three bluest star formation regions located outside the evolutionary tracks in 
Fig.~\ref{figure:nfig9} (bottom), we found that the mean age of the young 
stellar objects in Arm~A is $3.7\pm2.2$~Myr versus $6.0\pm1.1$~Myr for the 
star formation regions in Arm~B. Note that the star formation regions in 
Arm~A are younger than the complexes ($3.0\pm2.2$~Myr versus $4.7\pm1.9$~Myr).

\begin{figure}
\vspace{8.6mm}
\resizebox{0.87\hsize}{!}{\includegraphics[angle=000]{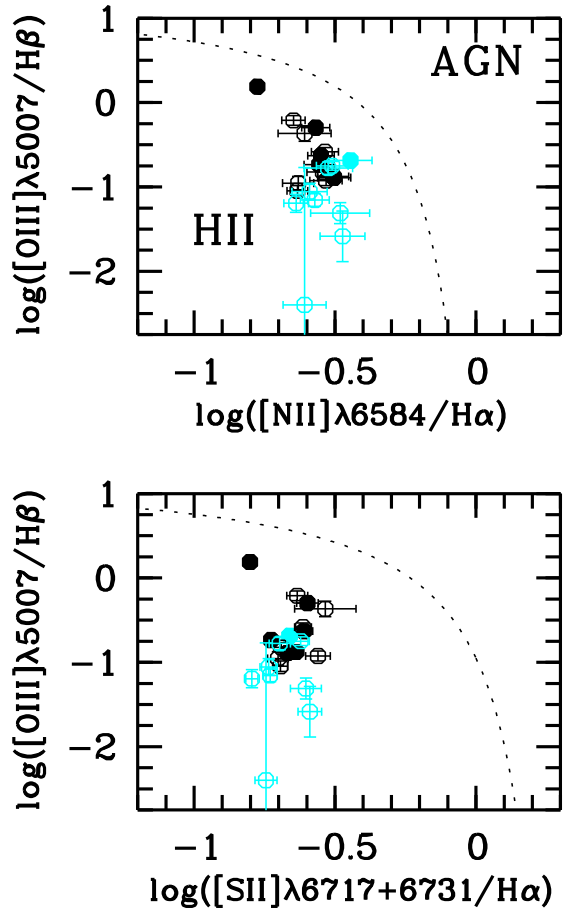}}
\caption{
Emission-line diagnostic diagrams for star formation regions in Arm~A 
and Arm~B. The curves represent upper boundaries for photoionized 
nebulae defined by \citet{kewley2006}. Symbols are the same as in 
Fig.~\ref{figure:nfig8}. 
}
\label{figure:fig_diag}
\end{figure}

\subsection{Chemical abundances of star formation regions}

\begin{figure*}
\vspace{1.0mm}
\resizebox{0.90\hsize}{!}{\includegraphics[angle=000]{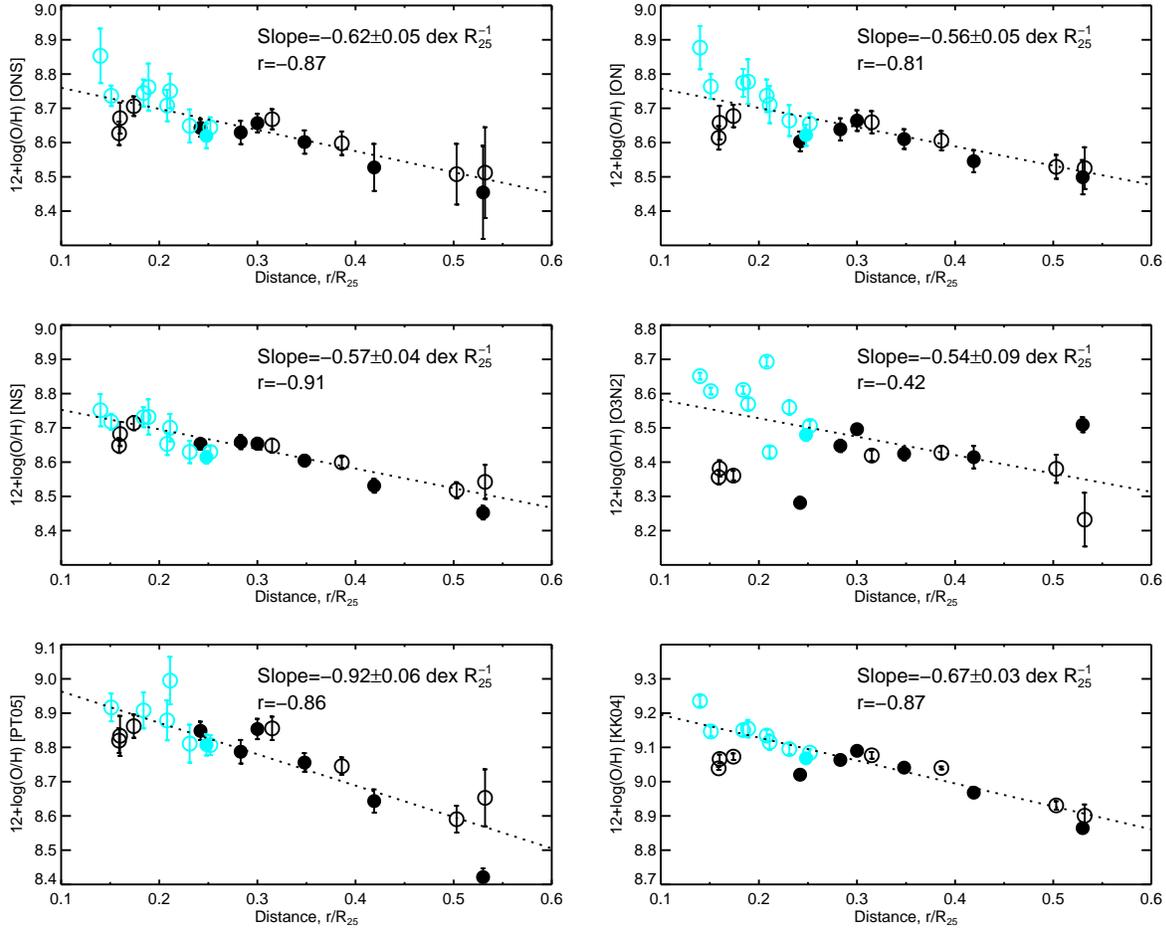}}
\caption{
Radial distribution of oxygen abundances in the galaxy. Oxygen abundances 
were obtained using six empirical methods (see text). Symbols are the same 
as in Fig.~\ref{figure:nfig8}. 
}
\label{figure:fig_oh_rad}
\end{figure*}

We selected a homogeneous sample of star formation regions, which have been 
studied using integral field spectroscopy techniques \citep{rosales2011}. 
The sample includes 22 regions out of 30 in both arms (see column (6) 
in Table~\ref{table:sfrs_id}).

The aim of this section is to compare the metallicities of the ionized 
gas of star formation regions located in different spiral arms. Oxygen is 
the most abundant heavy element in interstellar medium, so its abundance is 
the best indicator of gas metallicity.

It is useful to study the nitrogen-to-oxygen abundances ratio in galaxies 
for understanding their chemical evolution due to the difference of nature 
of these elements. Nitrogen is ejected into interstellar medium by both 
low- and intermediate-mass stars and massive stars, whereas oxygen is created 
only in the last ones. Analysis of O/N--O/H plane may allow us to 
arrive to conclusions about star formation rate and history of star-forming 
galaxies \citep{molla2006}.

The most accurate way to estimate the oxygen and nitrogen abundance is 
the so-called 'direct' temperature-based method. However, the direct 
method is unavailable for the objects, studied here, because of the 
absence of temperature-sensitive auroral lines, such as [O\,{\sc iii}] 
$\lambda4363$. We used several most popular empirical methods: ONS, ON 
\citep{pilyugin2010} and NS \citep{pilyugin2011}. Oxygen abundance have 
been estimated also by PT05 \citep{pilyugin2005}, O3N2 \citep{pettini2004} 
and KK04 \citep{kobul2004} empirical methods.

At this point, the following question arises: do methods calibrated on 
pure H\,{\sc ii} regions or on photoionization models give reliable 
estimations when applied to real star formation regions in NGC~628? 
To answer this question, we plotted the traditional 
[O\,{\sc iii}] $\lambda$5007/H$\beta$ versus 
[N\,{\sc ii}] $\lambda$6584/H$\alpha$ and 
[S\,{\sc ii}] $\lambda$6717,~6731/H$\alpha$ diagnostic diagrams for 
investigated regions in Arms~A and B in Fig.~\ref{figure:fig_diag}. 
Dashed lines denote upper boundaries for photoionized nebulae defined by 
\citet{kewley2006}. As one can see from this figure, all regions lie 
within the photoionization area and do not show signatures of shock 
excitation. This indicates that the empirical methods used are reliable.

Another feature that is clearly seen from Fig.~\ref{figure:fig_diag} is 
that complexes from Arm~B have lower [O\,{\sc iii}]/H$\beta$ values than 
those from Arm~A for the same ratio of [S\,{\sc ii}]/H$\alpha$ and 
[N\,{\sc ii}]/H$\alpha$. This can be easily explained by the lower 
ionisation parameter in regions from Arm~B \citep[see, for example, 
photoionisation models constructed by][]{levesque2010}.

\begin{figure*}
\vspace{1.0mm}
\resizebox{0.90\hsize}{!}{\includegraphics[angle=000]{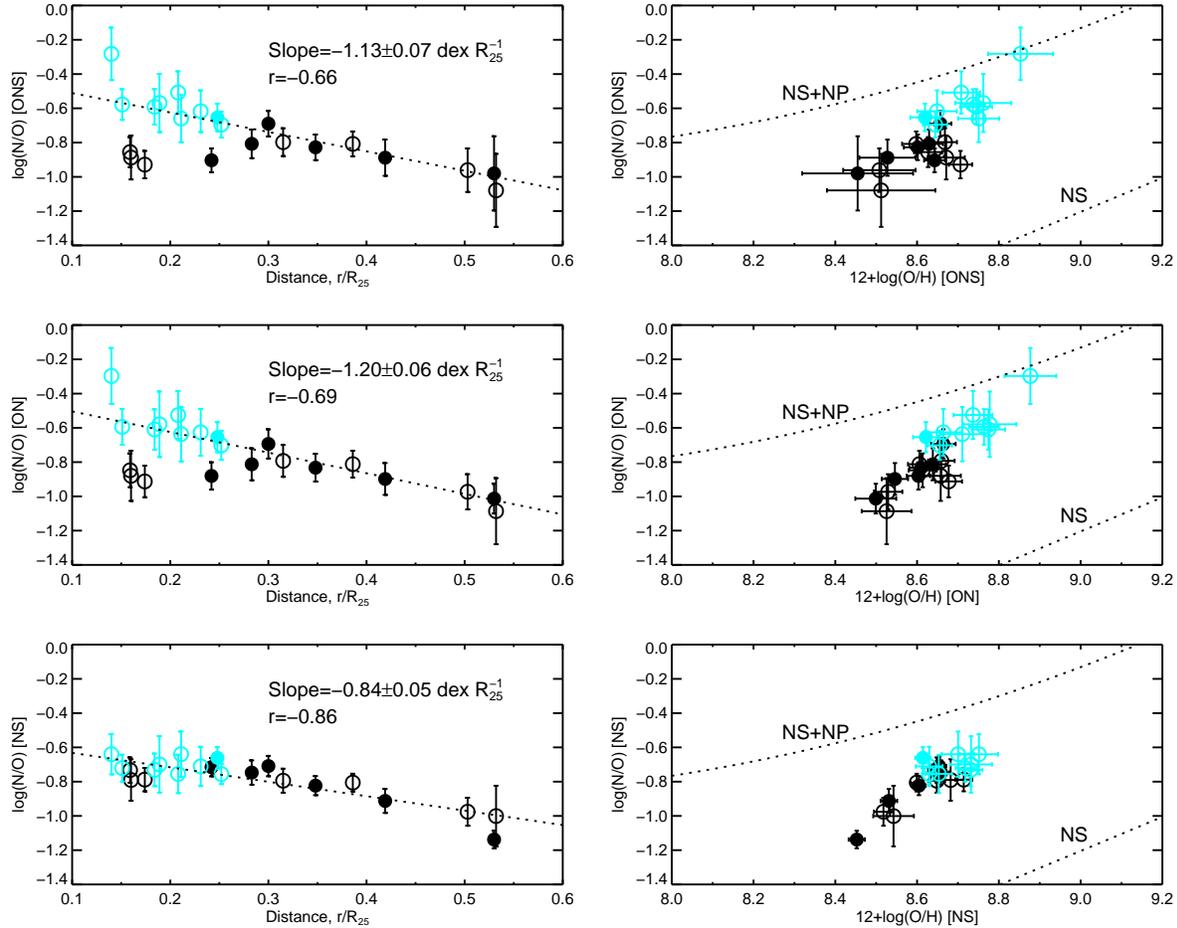}}
\caption{
N/O gradient of investigated star formation regions with their distance 
from the centre (left) and with oxygen abundance O/H (right). Abundances 
were obtained using three empirical methods (see text). Dashed lines on the 
right panels show possible boundaries for data points on the N/O--O/H 
plane under the assumption of close box model for secondary (NS) and both 
primary and secondary (NS+NP) nitrogen. Symbols are the same as in 
Fig.~\ref{figure:nfig8}. 
}
\label{figure:fig_no_oh}
\end{figure*}

In recent years, several authors performed detailed comparisons of 
abundance estimation methods and found their discrepancies 
\citep[see e.g.][and references therein]{kewley2008}. \citet{lopez2012} 
analysed model spectra of H\,{\sc ii} regions and showed that theoretical 
methods, such as KK04, give overestimated values of oxygen abundance in 
comparison with 'direct' $T_e$ method, whereas empirical ON, NS and ONS 
methods are in good agreement. Investigations of individual H\,{\sc ii} 
regions in nearby galaxies confirm that result 
\citep[see e.g.][]{egorov2012}. The situation is similar for star formation 
complexes in NGC~628 (see Fig.~\ref{figure:fig_oh_rad}). Oxygen abundances 
obtained by KK04 and O3N2 are slightly higher than those obtained by PT05, 
ON, ONS and NS methods. Possibly, this fact may be due to the large 
size of investigated regions, where local temperature inhomogeneities play 
an important role and have to be taken into account.

The oxygen abundance distribution along the radius of the galaxy 
shows a significant gradient. It was studied by \citet{rosales2011} 
using 4 methods of abundance determination, including O3N2 and KK04. 
We estimated oxygen abundance gradient by linear $\chi^2$ fitting of data 
points obtained with six empirical methods. The results are shown in 
Fig.~\ref{figure:fig_oh_rad}. The absolute value of the correlation 
coefficient, $r$, for almost all dependences shown in 
Fig.~\ref{figure:fig_oh_rad} is greater than 0.8 that corresponds to a 
fine linear approximation. There is only one exception -- the O3N2 
abundance versus distance where $r=-0.42$ and abundance measurements 
show a wide spread. It is not surprising because the accuracy of the 
O3N2 method is lower than of other applied methods (about 0.2~dex in 
comparison with 0.1~dex for other). The values of the gradient obtained 
are in good agreement for the ONS, ON, NS and PT05 methods, slightly higher 
for KK04 and much higher for O3N2 methods. Note that the slope of O/H 
dependence on radius obtained by KK04 method is in good agreement with 
\citet{rosales2011} estimations. This is not surprising because we used their 
reported fluxes. But our gradient, that we obtained with O3N2 method, is much 
steeper than the one reported by \citet{rosales2011}. This may be caused by 
using only a small sample of their data points.

Fig.~\ref{figure:fig_no_oh} shows the N/O ratios as a function of the 
distance from the galaxy centre (left-hand panels) and the oxygen abundances 
(right-hand panels) obtained with the ONS, ON and NS methods. 
Note that the N/O ratios with the galactocentric radius shown in 
Fig.~\ref{figure:fig_no_oh} are in good agreement with the results of 
\citet{berg2013}, who found the extrapolated central N/O ratio 
$-0.45\pm0.08$~dex and the slope of the N/O ratio gradient 
$-1.10\pm0.14$~dex\,$R_{25}^{-1}$ within the optical radius, $R_{25}$. 
Analysis of these dependences may be of help in answering the 
question on the nature of the nitrogen in the star formation complexes 
under study. If nitrogen is mostly primary (NP), then N/O ratio should be 
constant, but if it is secondary, the N/O ratio grows with oxygen 
abundance increase. Fig.~\ref{figure:fig_no_oh} shows exactly the same 
linear dependence. Moreover, the variation of the N/O gradient with 
the galactocentric radius is steeper than for $12+\log(\mathrm{O/H})$. 
This can be interpreted as evidence of the predominantly secondary 
nature of nitrogen in the star formation complexes under study. The trend in 
the evolution of the ratio N/O with $12+\log(\mathrm{O/H})$ shown in 
Fig.~\ref{figure:fig_no_oh} (right-hand panels) is in good agreement with 
those found in other Sc-type galaxies \citep{vilacostas1993}.

Fig.~\ref{figure:fig_no_oh} shows a clear separation between the 
properties of the star formation regions hosted by Arm~A and Arm~B. The 
regions from Arm~B show higher N/O ratio at a higher oxygen abundance. It was 
shown recently \citep[see e.g.][]{molla2010,mallery2007} that the location of 
a region in the N/O--O/H plane is related to the specific star formation 
rate, SFR, per unit mass in stars (sSFR). The higher values of N/O correspond 
to the smaller sSFR. If the sSFR is small, star formation could 
have been high in the past, at the earlier times of evolution. The gas 
was consumed and therefore the SFR decreased and is now small. And conversely, 
when the efficiency to form stars is low, the star formation rate 
increases over time and the present SFR is high. So, the N/O--O/H planes 
in Fig.~\ref{figure:fig_no_oh} may be explained if we propose that complexes 
in Arm~B had a higher SFR in the past, but now it is lower than for 
Arm~A. As we will see further, that is possibly our case (see 
Fig.~\ref{figure:nfig23}).

There are several regions from Arm~A and Arm~B that have similar oxygen 
abundance, but different N/O ratios. All three methods, used to estimate 
these values, give similar results -- complexes from Arm~B have slightly 
higher N/O ratios for a given $12+\log(\mathrm{O/H})$. This may be 
easily explained if regions from Arm~A are younger than those from Arm~B. 
In that case nitrogen could not enrich the interstellar medium in Arm~A 
because of the delay of the nitrogen appearance in the interstellar 
medium with respect to oxygen. This is supported by the results of 
\citet{sonbas2010}. Their search for supernova remnants in NGC~628 gave nine 
SNR candidates, five of them in Arm~A. Two out of three latest supernovae are 
also located in Arm~A (see Fig.~\ref{figure:nfig5b}).

\subsection{Star formation region luminosity function}

The distribution of star formation regions by mass, as well as an upper 
limit for the mass of these regions, depends on properties of 
interstellar medium such as gas density and pressure and correlates with the 
overall star formation rate 
\citep{elmegreen1997,kennicutt1998,billett2002,larsen2002}. Nevertheless, 
most studies of properties of star formation region populations have 
focused on the model-independent luminosity function 
\citep[see e.g.][]{haas2008,mora2009}. 

In order to further compare the properties of the star formation regions 
in Arms~A and B, we have constructed the luminosity function for the 
brightest relevant objects in both arms. In contrast to \citet{larsen2002}, 
\citet{haas2008} and \citet{mora2009}, we used ultraviolet luminosities, as 
they are most sensitive to the presence of young stellar 
populations. A standard power-law luminosity function of the 
form 
\begin{equation}
dN(L_{\rm FUV})/dL_{\rm FUV} = \beta L_{\rm FUV}^{\alpha}
\label{equation:lf}
\end{equation}
was adopted. It was converted to the form 
\begin{equation}
\log N = a{\rm FUV}+b
\label{equation:lf2}
\end{equation}
for the fitting, where the variables $\alpha$, $\beta$ in 
Eq.~(\ref{equation:lf}) and $a$, $b$ in Eq.~(\ref{equation:lf2}) are related 
as $\alpha = -2.5a-1$ and $\beta = 2.5(\ln 10)^{-1}10^{b+4.8a}$, 
respectively.

\begin{figure}
\vspace{7.2mm}
\resizebox{0.90\hsize}{!}{\includegraphics[angle=000]{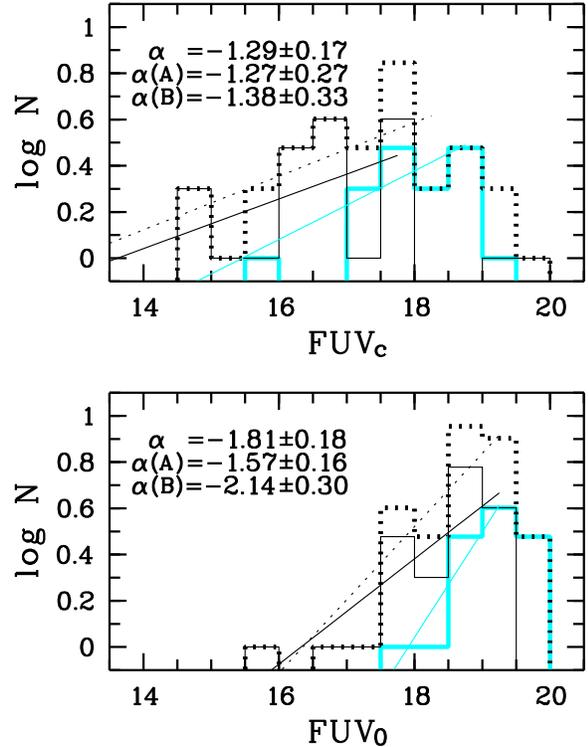}}
\caption{Luminosity functions for the regions using corrected for internal 
absorption (top panel) and uncorrected FUV magnitudes (bottom panel). 
Thick dotted histograms are luminosity functions for the regions in both 
arms, thin black solid histograms are functions for the regions in 
Arm~A, and thick grey histograms are functions for the regions in
Arm~B. Black dotted, black solid, and grey solid straight lines 
represent the power-law fit of the form of Eq.~(\ref{equation:lf}) for the 
samples of star formation regions in both arms, Arm~A and Arm~B, 
respectively.
}
\label{figure:nfig12n}
\end{figure}

The constructed star formation region luminosity functions are shown in 
Fig.~\ref{figure:nfig12n}. Each histogram was fitted using the normal 
least-squares method to an expression of the form of 
Eq.~(\ref{equation:lf2}). The results of the fitting are summarised in 
Table~\ref{table:lf_fit}.

Usually, researchers of cluster luminosity functions obtain internal 
extinction coefficients from evolutionary synthesis models 
\citep{larsen2002,mora2009}. The luminosity functions of the star 
formation regions have been obtained using both corrected and uncorrected 
(for interstellar absorption) magnitudes. The interstellar absorption 
coefficients are model-independent. We can only provide a rough estimate of 
the slope of the luminosity function for the brightest star formation regions 
in the spiral arms because of the small number statistics.

\begin{table}
\caption[]{\label{table:lf_fit}
Luminosity function coefficients.
}
\begin{center}
\begin{tabular}{ccccc} \hline \hline
Arm  & $\alpha$       & $a$ & $b$ & Fit interval \\
\hline
A   & -1.27$\pm$0.27 & 0.11$\pm$0.11 & -1.46$\pm$1.74 & FUV$_{\rm c} < 18.0$ \\
B   & -1.38$\pm$0.33 & 0.15$\pm$0.13 & -2.30$\pm$1.50 & FUV$_{\rm c} < 19.0$ \\
A+B & -1.29$\pm$0.17 & 0.12$\pm$0.07 & -1.50$\pm$1.14 & FUV$_{\rm c} < 18.5$ \\
\hline
A   & -1.57$\pm$0.16 & 0.23$\pm$0.06 & -3.72$\pm$1.10 & FUV$_0 < 19.5$ \\
B   & -2.14$\pm$0.30 & 0.46$\pm$0.12 & -8.18$\pm$2.29 & FUV$_0 < 19.5$ \\
A+B & -1.81$\pm$0.18 & 0.32$\pm$0.07 & -5.17$\pm$1.24 & FUV$_0 < 19.5$ \\
\hline
\end{tabular}
\end{center}
\end{table}

Obtained slopes of luminosity function, based on an uncorrected FUV data, 
differ for the region population in Arm~A and B (Fig.~\ref{figure:nfig12n}). 
The slope for the region population in Arm~B is typical for brightest young 
cluster populations in galaxies \citep{zepf1999,whitmore1999,larsen2002,
dolphin2002,grijs2003,gieles2006,haas2008,mora2009}. A more gently sloping 
function is obtained for the star formation region population in Arm~A. 
A value of the slope $\alpha = -1.5$ is close to the results of 
\citet{bergh1984} for the Milky Way open clusters, \citet{whitmore1999} and 
\citet{haas2008} for faint clusters in the Antennae and M51, respectively. 
The united population of star formation regions have an intermediate 
slope of luminosity function (Fig.~\ref{figure:nfig12n}).

A surprising result was obtained for the star formation region 
luminosity function when we used corrected FUV magnitudes. The same, within 
errors, shallow slope is found for the star formation regions population 
in both spiral arms. The flat distribution can be a result of selection; 
we lost objects with high extinction, which are slightly fainter than 
19.7~mag in FUV. However, this effect must be the same for the star 
formation regions populations in both arms. Note that the large error in 
the slope for the star formation regions sample in Arm~B is due to the 
large uncertainty of corrected FUV magnitude of the brightest star 
formation region b11 (see Table~\ref{table:sfrs_photcor}).

\subsection{Sizes and size distribution functions of star formation 
regions}

\begin{figure}
\vspace{3.0mm}
\resizebox{1.00\hsize}{!}{\includegraphics[angle=000]{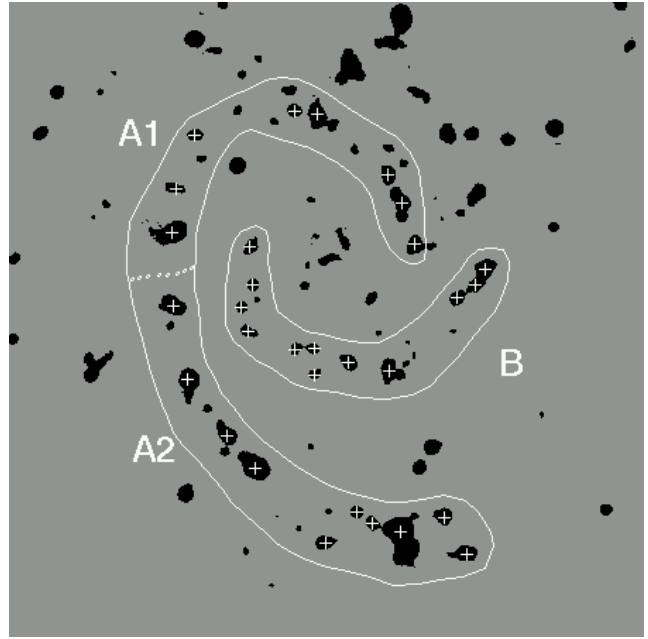}}
\caption{Map of star formation regions in NGC~628. Areas with a surface 
brightness $\mu({\rm FUV}_0) < 23.63$~mag\,arcsec$^{-2}$ are indicated by 
black colour. White curves show boundaries of Arms~A and B. The dotted line 
shows the boundary between inner (A1) and outer (A2) parts of Arm~A. 
The white crosses indicate positions of the regions from 
Table~\ref{table:sfrs_id}. The size of the image is $6.0\times6.0$ 
arcmin$^2$. North is upward and east is to the left.
}
\label{figure:nfig20}
\end{figure}

To measure sizes of the star formation regions, we used the following 
technique: (i) the mean intensity level of the background in FUV, 
$\langle F \rangle$, and its standard deviation, $\sigma$, within the arms but 
outside the star formation regions were found, (ii) the cutoff intensity, 
$F_{\rm cut} = \langle F \rangle + 5\sigma$ was calculated, (iii) all pixels 
in the FUV image with the intensity $F > F_{\rm cut}$ were selected. 
The cutoff intensity $F_{\rm cut}$ corresponds to a surface brightness 
$\mu({\rm FUV}_0) = 23.68\pm0.10$~mag\,arcsec$^{-2}$. Areas within Arms~A 
and B with a surface brightness level higher than 23.63~mag\,arcsec$^{-2}$ in 
FUV were identified and measured (Fig.~\ref{figure:nfig20}). We found 56 
regions in total. Characteristic diameters $d$ of star formation regions were 
defined as
\begin{equation}
d = 2\sqrt{S/\pi}, \nonumber
\label{equation:sizes}
\end{equation}
where $S$ is the area of selected regions. Diameters of star formation 
regions from our sample are given in the last column of 
Table~\ref{table:sfrs_photcor}. Errors in determining the diameters of the 
objects are caused by the accuracy of determining the value of $F_{\rm cut}$.

Arm~A is twice as long as Arm~B. To compare the size distribution of 
the regions in Arms~A and B on the same galactocentic distance range, 
we divided Arm~A into inner (A1) and outer (A2) parts. The end of inner 
part of Arm~A corresponds to the end of Arm~B (Fig.~\ref{figure:nfig20}). 
It looks like the inner part of Arm~A (A1) and Arm~B in 
Fig.~\ref{figure:nfig20} are 'classic' spiral arms -- as regards to their 
inner structure they are similar and seems to be that both arms 
show the same age (composition) gradient across the arm. They also have 
approximately the same length. 

The characteristic diameters of 30 star formation regions from our 
samples are in the range 225--800~pc (Table~\ref{table:sfrs_photcor}), 
the diameters of the other 26 star formation regions are smaller, 
30~pc $< d <$ 250~pc.

The three brightest star formation complexes in Arm~A (A8, A12 and A16) 
have characteristic diameters $d>500$~pc. All these complexes are double in 
reality, as seen in $U$ and H$\alpha$ images of the galaxy 
(Figs.~\ref{figure:nfig5b}, \ref{figure:nfig20}). The size of the largest 
complex in Arm~B does not exceed 450~pc (Table~\ref{table:sfrs_photcor}).

Among the 30 brightest star formation regions, the regions in Arm~A are 
larger than the ones in Arm~B. Moreover, the mean diameter of the 
regions in the inner part of Arm~A is slightly larger than the mean diameter 
of the objects in Arm~B (Table~\ref{table:sfrs_size}).

The number of star formation regions in the arms as a whole, Arm~A, the 
inner part of Arm~A, and Arm~B decreases with the growth of diameter. 
The distribution of the regions in the outer part of Arm~A is flat until 
$\sim 400-500$~pc (Table~\ref{table:sfrs_size}).

Detailed exploration of the size distribution of objects in NGC~628 was 
made in \citet{elmegreen2006} in the range of scales from 2 to 
110~pc\footnote{For an adopted distance of 7.2~Mpc.}, based 
on {\it HST} images. \citet{elmegreen2006} found that the cumulative size
distribution follows a power law, with slope $\gamma \approx -1.5$. 
The closest value of the slope of the cumulative size distribution 
function was found for the star formation regions from the list of 
\citet{ivanov1992} that is in the range 30--110~pc. The size distribution of 
larger objects, H\,{\sc ii} regions studied by \citet{hodge1976}, satisfies 
a power law with slope $\gamma \approx -3.5$ in the range 100--300~pc. The size distribution 
of complexes of \citet{ivanov1992} gives $\gamma = -4.1$ in the range 
500--1000~pc.

\begin{figure}
\vspace{7.0mm}
\resizebox{0.85\hsize}{!}{\includegraphics[angle=000]{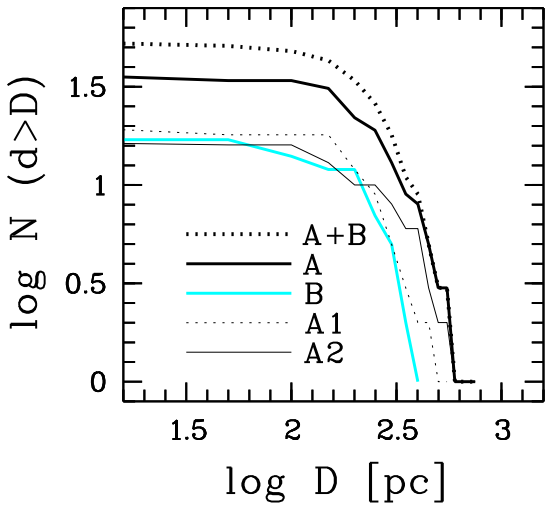}}
\caption{Cumulative size distribution function for the regions in 
both spiral arms (thick dotted line), Arm~A (thick solid line), Arm~B 
(thick grey solid line), the inner part of Arm~A (thin dotted line), 
and the outer part of Arm~A (thin solid line).
}
\label{figure:nfig22}
\end{figure}

Following \citet{elmegreen2006}, we constructed the cumulative size 
distribution function for star formation regions in the spiral arms of 
NGC~628 in the form 
\begin{eqnarray}
N (d>D) \propto D^{\gamma}, \nonumber
\end{eqnarray}
where $N$ is the integrated number of objects that have a diameter $d$ 
greater than some diameter $D$ (Fig.~\ref{figure:nfig22}).

The slope of the power law for the size distribution is approximately the 
same for star formation region populations in both arms as a whole, 
Arms~A and B, and the inner part of Arm~A for a size range of 200 to 
400~pc, $\gamma \approx -2$ (Table~\ref{table:sfrs_size}). The 
distribution of the largest regions in both arms as a whole, and Arm~A and 
Arm~B separately, satisfies a power law with slope $\gamma \approx -5$. 
Differences in the size distribution are found between the star formation 
region populations of the inner and outer parts of Arm~A, and between the 
populations of Arm~B and the inner part of Arm~A (Fig.~\ref{figure:nfig22}, 
Table~\ref{table:sfrs_size}).

\begin{table}
\caption[]{\label{table:sfrs_size}
Mean diameters and size function coefficients.
}
\begin{center}
\begin{tabular}{cccccc} \hline \hline
Arm  & $\langle d \rangle^a$ & $\gamma$ & Range & $\gamma$ & Range \\
     & (pc)                  &          & (pc)  &          & (pc)  \\
\hline
A   & $400\pm135$ & -1.6 & 200-400 & -4.5 & 400-650 \\
A1  & $360\pm100$ & -2.1 & 200-300 & -2.7 & 300-600 \\
A2  & $435\pm160$ & -0.8 & 200-400 & -3.6 & 400-650 \\
B   & $295\pm70$  & -2.2 & 200-300 & -5.6 & 300-450 \\
A+B & $360\pm125$ & -2.0 & 200-400 & -4.7 & 400-650 \\
\hline
\end{tabular}
\end{center}
\begin{flushleft}
$^a$ The mean diameters.
\end{flushleft}
\end{table}

The size distribution of the star formation region population in Arm~B 
repeats the distribution of the region samples in Arm~A with a displacement 
$\log D \approx 0.2$ (Fig.~\ref{figure:nfig22}). The size distribution 
function of the star formation region population in the inner part of Arm~A 
have approximately the same slope in the entire range studied here 
(Table~\ref{table:sfrs_size}). The size function curves for the populations 
of Arm~B and the inner part of Arm~A are very close in the range from 200 to 
300~pc, but they vary considerably in the range from 300 to 500~pc. The size 
distribution function of the star formation regions sample in the outer part 
of Arm~A is characterized by the shallow slope, $\gamma = -0.8$, in the 
intermediate range from 200 to 400~pc (Table~\ref{table:sfrs_size}).

\subsection{Star formation rates within star formation regions}

As we pointed out above, the distributions of star formation regions 
by mass and luminosity and the upper limits of mass and size of regions 
correlate with the overall star formation rate and depend on properties of 
interstellar medium. We measured the star formation rates, SFR, and the 
surface densities of star formation rate, $\Sigma_{\rm SFR}$, within the star 
formation regions using obtained FUV magnitudes, H$\alpha$ luminosities and 
sizes. To accomplish this, we adopt the conversion factor of FUV luminosity 
to star formation rate of \citet{iglesias2006}, namely
\begin{eqnarray}
{\rm SFR (}M_\odot \,{\rm yr^{-1})} = 8.13\times10^{-44} L_{\rm FUV} {\rm (erg\,s^{-1})}, \nonumber
\end{eqnarray}
in the form
\begin{equation}
{\rm SFR (}M_\odot \,{\rm yr^{-1})} = 7.0\times10^{-8}\times10^{-0.4M({\rm FUV})_{\rm c}},
\label{equation:sfr_fuv}
\end{equation}
and the conversion factor of H$\alpha$ luminosity to star formation rate of 
\citet{kennicutt1998b}:
\begin{equation}
{\rm SFR (}M_\odot \,{\rm yr^{-1})} = 7.9\times10^{-42} L_{{\rm H}\alpha} {\rm (erg\,s^{-1})}.
\label{equation:sfr_ha}
\end{equation}

\begin{figure}
\vspace{3.5mm}
\resizebox{1.00\hsize}{!}{\includegraphics[angle=000]{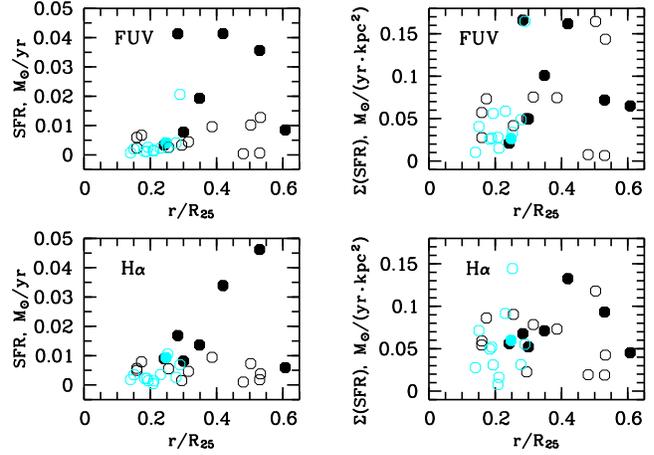}}
\caption{
Radial distributions of star formation rate (left panels) and the surface 
density of star formation rate (right panels) within regions based on 
their luminosities in FUV (top panels) and H$\alpha$ (bottom panels). Other 
symbols are the same as in Fig.~\ref{figure:nfig8}.
}
\label{figure:nfig23}
\end{figure}

The surface densities of star formation rate within the star formation 
regions are measured as
\begin{eqnarray}
\Sigma_{\rm SFR} = {\rm SFR}/S, \nonumber
\end{eqnarray}
where SFR and $S$ are obtained from Eqs.~(\ref{equation:sizes}), 
(\ref{equation:sfr_fuv}) and (\ref{equation:sfr_ha}).

Note that the total star formation rate within studied regions, 
$\approx 0.25~M_\odot \,{\rm yr^{-1}}$, is one third of the full SFR in 
NGC~628 by data of \citet{calzetti2010}, who estimated 
${\rm SFR} = 0.7\pm0.2~M_\odot \,{\rm yr^{-1}}$ in the galaxy as a whole.

Densities of SFR within complexes are typical for star formation regions 
and comparable to results of \citet{bastian2005}, who found 
$\Sigma_{\rm SFR} = 0.06-0.07~M_\odot \,{\rm yr^{-1} \, kpc^{-2}}$ for 
ordinary complexes in M~51. Bright complexes and fainter star formation 
regions have similar surface densities of star formation rate 
(Fig.~\ref{figure:nfig23}).

\begin{figure}
\vspace{7.0mm}
\resizebox{0.87\hsize}{!}{\includegraphics[angle=000]{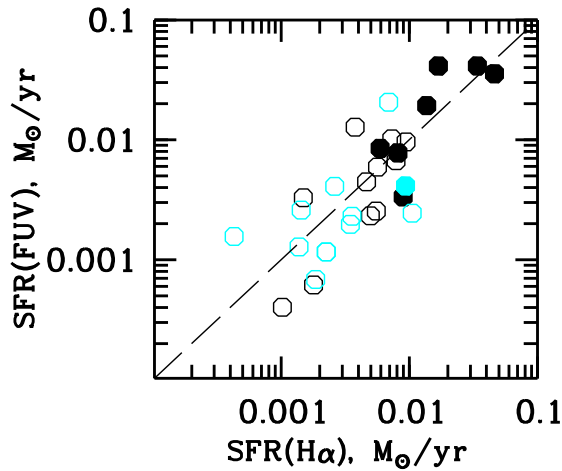}}
\caption{
Comparison between SFR(FUV) and SFR(H$\alpha$) of star formation regions.
Other symbols are the same as in Fig.~\ref{figure:nfig8}.
}
\label{figure:nfig23b}
\end{figure}

In spite of the difference in the estimation of SFR(FUV) and 
SFR(H$\alpha$) for some regions, the results are in agreement with each 
other, in general, 
$\langle \Sigma_{\rm SFR}({\rm FUV}) \rangle = 0.063~ 
M_\odot \,{\rm yr^{-1} \, kpc^{-2}}$ and 
$\langle \Sigma_{\rm SFR}({\rm H}\alpha) \rangle = 0.061~ 
M_\odot \,{\rm yr^{-1} \, kpc^{-2}}$. Comparison between SFR(FUV) and 
SFR(H$\alpha$) of the star formation regions is presented in 
Fig.~\ref{figure:nfig23b}.

A dependence of SFR density on galactocentic distance is found for 
objects of Arm~A: $\langle \Sigma_{\rm SFR} \rangle \approx 0.07~ 
M_\odot \,{\rm yr^{-1} \, kpc^{-2}}$ for all regions in Arm~A, and only 
$\approx 0.06~M_\odot \,{\rm yr^{-1} \, kpc^{-2}}$ for regions with 
$r/R_{25} < 0.32$. The regions in Arm~B have a smaller surface density of 
SFR, on average, than the objects in Arm~A, 
$\langle \Sigma_{\rm SFR}\rangle \approx 0.05~ 
M_\odot \,{\rm yr^{-1} \, kpc^{-2}}$.

Differences in the density of star formation rate within star formation 
regions in Arms~A and B may indicate differences in interstellar 
medium parameters between the arms.

\section{Discussion}

Spiral density waves can play an important role in asymmetric star 
formation in spiral arms. \citet{henry2003} showed on the example of the 
asymmetry in the spiral arms of M51, that the variable star formation can be 
caused by more than one spiral density wave. Moreover, an asymmetry in 
the spiral arms of NGC~628 has been detected in the observed two-dimensional 
field of radial velocities of the gas in the disc of the galaxy 
\citep{sakhibov2004}. Fourier analysis of the azimuthal distribution of 
the observed radial velocities in annular (ring) zones at different distances 
from the centre of the disc shows the existence of two spiral density waves 
\citep[see Fig.~1a in][]{sakhibov2004}, the one-armed wave in addition to the 
dominant two-armed one. This additional spiral density wave corresponds to 
the star formation asymmetry in the two main symmetrical arms revealed 
through the computer-enhanced images of the galaxies by \citet{elmegreen1992}. 
In the case of NGC~628, the relatively lower SFR in Arm~B can be caused 
by the asymmetry of the spiral density waves in the galaxy.

In \citetalias{gusev2013a} we assumed that the drastic differences 
observed between the inner structures located in the spiral arms of 
NGC~628, one of which hosts the regular chain of large star complexes 
whereas the other does not, were the result of the existence of a 
regular magnetic field and the absence of the signature of a shock wave along 
Arm~A. Unfortunately, there are no appropriate magnetic field 
data for the studied part of NGC~628. The only data concerning the magnetic 
field were obtained by \citet{heald2009} who detected polarized emission at 
18 and 22~cm wavelengths from the outer part of the galaxy; their linear beam 
size was $1.9\times0.5$~kpc.

The hypothesis proposed in \citetalias{gusev2013a} is not the only 
possible one. Alternatively, asymmetries in spiral galaxies can be the 
result of gravitational interactions with another galaxy or galaxies at some 
point in their history. NGC~628 is a member of a small group of galaxies and 
its present state may well be the result of close encounters within the 
group. Close encounters may also trigger episodes of star formation. The 
same tidal forces that can deform the galaxy may also disrupt giant 
molecular clouds within the galaxy and induce their gravitational collapse. 
The numerical simulations of \citet{bottema2003} show that unbarred grand 
design galaxies, such as NGC~628, can only be generated by tidal forces 
resulting from an encounter with other galaxy. However, we believe that tidal 
interactions could not play a role in the origin of the observed 
asymmetrical pattern of star formation. It is well established that NGC~628 
cannot have undergone any encounter with satellites or other galaxies in the 
past 1~Gyr \citep{wakker1991,kamphuis1992}. The spiral filaments are possibly 
disturbed by interaction with the two large high velocity gas clouds on either 
side of the disc \citep{kamphuis1992,beckman2003}. However, these high 
velocity gas clouds are located symmetrically with respect to the centre 
\citep{kamphuis1992}. Residual velocity fields of both neutral and ionized 
gas show the absence of significant velocity deviations from the radial 
velocity \citep{kamphuis1992,fathi2007}.

In this paper we have found differences in photometric parameters and chemical 
abundance between the star formation regions in Arms~A and B. We suggest that 
these differences are the result of significant differences in the physical 
properties of interstellar medium in the opposite arms of NGC~628.

As is known, such physical processes as gravitational collapse and 
turbulence compression play a key role in creation and evolution of star 
formation regions over the wide range of scales, from smallest OB 
associations to largest star complexes 
\citep{efremov1995,elmegreen2000,elmegreen2006,elmegreen2002,elmegreen2006a}. 
The age range of stars within ordinary star formation regions is usually 
quite small ($\le10-15$~Myr) suggesting a coherent star formation mechanism, 
it separates them from large star complexes which have a much larger 
intrinsic age spread \citep{efremov1995}. This is well illustrated for 
the young stellar objects in Arm~A, where the large complexes are older than 
the star formation regions ($4.7\pm1.9$~Myr versus $3.0\pm2.2$~Myr). We 
suggest that the difference between photometric ages of star formation 
regions in Arms~A and B is a result of different star formation histories in 
them. The generation of shock-waves is the source of high pressure in Arm~B 
and probably within the inner part of Arm~A (A1). High pressure stimulates 
formation of dense star formation regions with an active star formation, 
including formation of massive ($M>10M_\odot$) stars \citep{billett2002}. 
High pressure from formed H\,{\sc ii} regions destroy molecular cloud 
cores \citep{elmegreen1983a}. As a result, SFR along Arm~B falls for several 
Myr, star formation regions here do not reach large mass/size, and they 
have approximately the same photometric ages, $6.0\pm1.1$~Myr.

The opposite case is observed in Arm~A. Pressure along Arm~A is lower 
than along Arm~B. As a result, 'doughy' large complexes are formed here. 
Initial SFR is low along Arm~A; massive stars are not formed immediately. 
The pressure increase driven by powerful stellar winds from the most 
massive stars is not high enough to destroy the largest cloud cores. As a 
result, the young stellar objects here have younger ages with a larger 
dispersion than the star formation regions in Arm~B ($3.7\pm2.2$~Myr versus 
$6.0\pm1.1$~Myr).

Thus, larger star complexes with a lower SFR in the past and a higher SFR now 
are observed in Arm~A, and smaller, more evolved star formation regions are 
observed in Arm~B. This hypothesis is supported by both abundance and 
photometric data. It is also fully consistent with the findings by 
\citet{sonbas2010}, who found nine SNR candidates in NGC~628. Five of them 
are located in Arm~A. Two out of three latest supernovae, SN~2003gd and 
2013ej, are also located in Arm~A (Fig.~\ref{figure:nfig5b}). Neither SNRs 
nor supernovae were found in Arm~B.

Note that all five SN remnants and two supernovae are located in Arm~A 
between complex A10 found in this work and star formation region a13 
(Fig.~\ref{figure:nfig5b}). It is where the star formation complexes and 
regions with the highest SFRs are observed (see Fig.~~\ref{figure:nfig23}). 
It is worth noting that SNR~9 and SN~2003gd are located within 
10--15~arcsec from our star formation region a13 (Fig.~\ref{figure:nfig5b}). 
Supernova 2003gd has the normal Type II-P, its progenitor was a red 
supergiant with initial mass $6-12 M_\odot$ \citep{vandyk2003,hendry2005}. 
The life time of such stars is $\sim10-50$~Myr. This is consistent with 
sustained star formation activity during last several tens Myr in this part 
of the arm.

The same slopes of the luminosity and size distribution functions for the 
sets of star formation regions in Arms~A and B, and the same characteristic 
separations, $\Lambda \approx 400$~pc \citepalias{gusev2013a} of star 
formation regions in both spiral arms, which depend on fundamental 
parameters of the medium, show that large scale fundamental properties 
of interstellar medium and kinematics of the galaxy have no principal 
differences in Arms~A and B.

In spite of the difference between parameters of star formation regions 
in the central part of Arm~A and other parts of the arms, the large 
scale (2~kpc and more) density of young stellar population along Arms~A and B 
is the same. Masses of star complexes in the central part of Arm~A are 3--4 
times as much as masses of star formation regions in other parts of Arms~A 
and B (Fig.~\ref{figure:nfig9}). However, the characteristic separations of 
star complexes are also 2--4 times as much as separations of star formation 
regions \citepalias{gusev2013a}.

We assume that the regular chain of star complexes in Arm~A can be explained 
by presence of the regular magnetic field and absence of the shock wave 
along the arm only for objects A8--A12 (the first five H\,{\sc ii} complexes 
from the Elmegreens' chain). In the outer part of Arm~A, star formation 
regions are located more chaotically, the largest star formation complex, 
A16, is observed here. This complex is the largest and the brightest one in 
FUV$_0$ in NGC~628 (Table~~\ref{table:sfrs_photcor}). Parameters of the 
complex can be related to its location near the corotation radius 
$R_{\rm cor}$ (\citet{sakhibov2004} obtained $R_{\rm cor} \approx 7$~kpc or 
$0.65r/R_{25}$ based on a Fourier analysis of the spatial distribution of 
radial velocities of the gas in the disc of NGC~628).

\section{Conclusions}

Photometric properties, chemical abundances and sizes of the 30 brightest 
star formation regions in the two principal arms of NGC~628 were studied, 
based on the {\it GALEX} ultraviolet, optical $UBVRI$, and H$\alpha$ surface 
photometry data.

We found, that the star formation regions in one (longer) arm (Arm~A) 
of NGC~628 of are systematically brighter and larger than the regions 
the other (shorter) arm. However, both luminosity and size distribution 
functions have approximately the same slopes for the samples of star 
formation regions in both arms. The star formation regions of Arm~A have 
a higher density of star formation rate than the regions in Arm~B. 
The regions from Arm~B show higher N/O ratio at a higher oxygen abundance, 
but they have lower ultraviolet and H$\alpha$ luminosities. Results of 
stellar evolutionary synthesis show that the brightest regions in Arm~A are 
younger than the ones in Arm~B ($3.7\pm2.2$~Myr versus 
$6.0\pm1.1$~Myr). The star complexes in Arm~A are slightly older than the 
star formation regions ($4.7\pm1.9$~Myr versus $3.0\pm2.2$~Myr).

The results can be explained if we suggest that star formation regions in 
Arm~B had higher star formation rate in the past, but now it is lower, 
than for opposite Arm~A.

Our results demonstrate that there is a difference in the inner structures 
and parameters of the interstellar medium between the two principal 
spiral arms of NGC~628. In spite of close sizes and spacing of star formation 
regions in Arm~B and in the inner part of Arm~A, modern star formation 
histories in Arms~A and B differ. Young stars in the central part of Arm~A 
($r/R_{25} = 0.28 - 0.42$) are grouping into the large complexes 
($d > 450$~pc). Smaller star formation regions are absent here.

\section*{Acknowledgements}
We are extremely grateful to the anonymous referee for enormous amount of 
detailed comments, most of which were very useful and formulated precisely 
enough to be directly incorporated into the paper text. The authors would like 
to thank Yu.~N.~Efremov (SAI MSU) and A.~E.~Piskunov (Institute of Astronomy 
of Russian Academy of Sciences) for helpful discussions. The authors are 
grateful to E.~V.~Shimanovskaya (SAI MSU) for help with editing this paper. 
The authors acknowledge the usage of the HyperLeda data base 
(http://leda.univ-lyon1.fr), the NASA/IPAC Extragalactic Database 
(http://ned.ipac.caltech.edu), Barbara A. Miculski archive for space 
telescopes (http://galex.stsci.edu), and the Padova group online server CMD 
(http://stev.oapd.inaf.it). This study was supported in part by the Russian 
Foundation for Basic Research (project no. 12--02--00827).

\end{document}